\documentclass[aps, prb, twocolumn, longbibliography, superscriptaddress]{revtex4-2}

\usepackage{amsmath,amssymb,bm,graphicx}
\usepackage{dcolumn}
\usepackage{color}
\usepackage[utf8]{inputenc}
\usepackage[T1]{fontenc}
\usepackage[colorlinks=true,citecolor=blue]{hyperref}
\hypersetup{colorlinks=true,citecolor=blue,linkcolor=blue,urlcolor=blue}

\begin{document}

\title{Polarons and Exciton-Polarons in Two-Dimensional Polar Materials}

\author{V.~Shahnazaryan}
\email{vanikshahnazaryan@gmail.com}
\affiliation{Abrikosov Center for Theoretical Physics,  Dolgoprudnyi, Moscow Region 141701, Russia}

\author{A.~Kudlis}
\email{andrewkudlis@gmail.com}
\affiliation{Science Institute, University of Iceland, Dunhagi 3, IS-107, Reykjavik, Iceland}
\affiliation{Abrikosov Center for Theoretical Physics, Dolgoprudnyi, Moscow Region 141701, Russia}

\author{I.~V.~Tokatly}
\affiliation{Departamento de Polímeros y Materiales Avanzados: Física, Química y Tecnología, Universidad del País Vasco, Avenida Tolosa 72, E-20018 San Sebastián, Spain}
\affiliation{Donostia International Physics Center (DIPC), E-20018 Donostia-San Sebastián, Spain}
\affiliation{IKERBASQUE, Basque Foundation for Science, Plaza Euskadi 5, 48009 Bilbao, Spain}

\date{\today}

\begin{abstract}
We propose a macroscopic theory of optical phonons, Fr{\"o}hlich polarons, and exciton-polarons in two-dimensional (2D) polar crystalline monolayers. Our theory extends the classical macroscopic formulation of the electron-phonon problem in three-dimensional (3D) polar crystals to the new generation of 2D materials. Similarly to the 3D case, in our approach, the effective electron-phonon Hamiltonian is parametrized solely in terms of macroscopic experimentally accessible quantities -- 2D polarizabilities of the monolayer at low and high frequencies. We derive the dispersion of long wave length longitudinal optical (LO) phonons, which can be viewed as a 2D form of the Lyddane-Sachs-Teller relation, and study the formation of 2D Fr{\"o}hlich polarons by adopting the intermediate coupling approximation. Finally, we apply this approach to excitons in polar 2D crystals and derive an effective potential of the electron-hole interaction dressed by LO phonons. Due to a specific dispersion of LO phonons, polarons and exciton-polarons in 2D materials exhibit unique features not found in their 3D counterparts. As an illustration, the polaron and exciton-polaron binding energies are computed for a representative set of 2D polar crystals, demonstrating the interplay between dimensionality, polarizability of materials, and electron-phonon coupling.
\end{abstract}

\maketitle

Polarons, quasiparticles arising from the interaction of charge carriers with lattice vibrations, have played a pivotal role in understanding transport and optical properties of polar materials \cite{alexandrov2010advances}.
The original concept of Landau-Pekar polaron \cite{landau1933electron,pekar1946local,landau1948effective} corresponds to an electron that locally polarizes the ionic crystals, which reduces its energy and may lead to autolocalizaton in the adiabatic self-consistent potential well. An alternative and complementary field theoretical picture of the electron dressed by a cloud of longitudinal optical (LO) phonons is known as the Fr\"ohlich polaron \cite{frohlich1950xx,lee1953motion,feynman1955slow}.
Polaronic effects in bulk polar crystals such as alkali halides and oxides are well studied both theoretically and experimentally \cite{devreese2009frohlich,franchini2021polarons}, providing important benchmarks for strong-coupling electron-phonon physics.
Here recent advances 
include accurate diagrammatic Monte-Carlo (MC) results \cite{prokofev1998polaron,hahn2018diagrammatic}, and first-principle calculations \cite{giustino2017electron,sio2019polarons,sio2019ab}.

The impact of LO phonons on exciton states leading to the formation of exciton-polarons is widely studied in II-VI and III-V polar semiconductors \cite{haken1958theorie,mahanti1972effective,bajaj1974effect,pollmann1977effective,kane1978pollmann}.
Recently, there has been a growing interest in exciton-polaron states in hybrid perovskites \cite{baranowski2020excitons}.
Here, the exciton-polaron concept allows to correctly address the main features of optical spectra both in bulk \cite{soufiani2015polaronic,menendez2015nonhydrogenic,filip2021phonon,wu2021strong,masharin2022polaron,masharin2023room,baranowski2024polaronic}, 
and layered hybrid perovskites \cite{simbula2021polaron,tao2021momentarily,tao2022dynamic,duan20242d,lei2024persistent,dyksik2024polaron,biswas2024exciton}.

Despite rapid progress in the fabrication and study of two-dimensional (2D) materials \cite{novoselov20162d}, there are a handful of observations of polaronic effects in 2D. This includes hexagonal boron nitride (hBN) \cite{chen2018emergence}, MoS$_2$ monolayer \cite{kang2018holstein}, SrTiO$_3$ surface \cite{chen2015observation,wang2016tailoring}.
{Theoretically, polarons in 2D have been mostly studied in the context of surface polarons that appear due to interaction with phonons on surfaces of 3D polar crystals
\cite{Sak1972,xiaoguang1985exact,hahn2018diagrammatic,li2018optical}. Only recently, the problem of intrinsically 2D polarons has been addressed within a first principle extension of Landau-Pekar polaron theory \cite{sio2023polarons}, and perturbatively by computing
polaron correction to Landau levels in the presence of magnetic field \cite{chen2018magneto}.}
The exciton-polaron problem for a single layer GaN was addressed within {\em ab initio} many-body theory \cite{alvertis2024phonon,lee2024phonon} by perturbatively correcting the kernel in the Bethe-Salpeter equation.
{Very recently, quasi 2D exciton-polarons were studied with path integral MC \cite{rana2024interplay}, adopting a model electron-phonon coupling of Refs.~\cite{sohier2016two,sio2022unified}.}

\begin{figure}[b]
    \includegraphics[width=1\columnwidth]{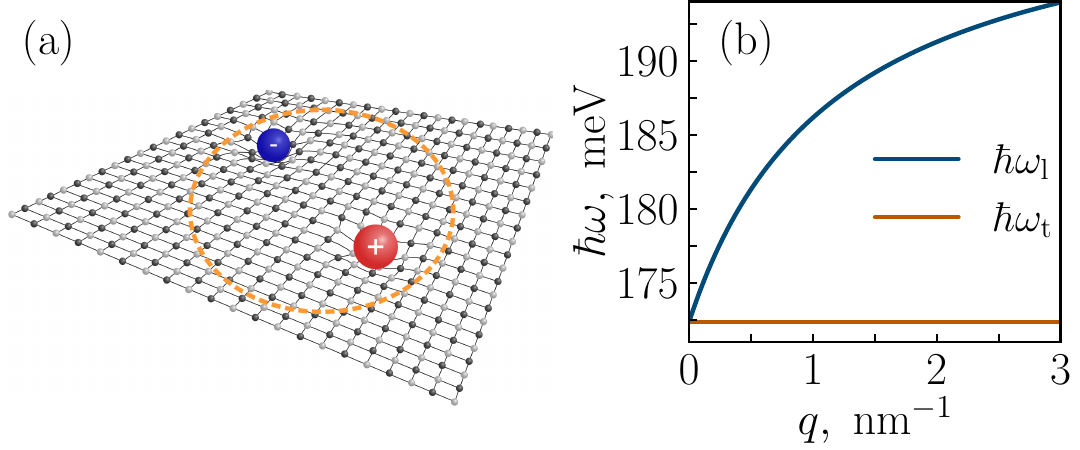}
    \caption{(a) Schematic picture of a 2D polar crystal layer hosting an electron-hole pair dressed by LO phonons, leading to the formation of  exciton-polaron. 
    (b) Dispersion of LO and TO phonon modes for a hBN monolayer. 
    }
    \label{fig:1}
\end{figure}

In 2D insulators, a reduced and highly nonlocal dielectric screening \cite{Keldysh1979,Rytova1967,cudazzo2011dielectric} results in a very unusual dispersion of LO phonons \cite{sohier2016two,sohier2017breakdown}. Apparently, the correct treatment of the specific 2D screening is crucial for all phenomena related to LO phonons and electron-phonon interaction in polar 2D systems. Some aspects of this problem have been studied and understood using a thin dielectric slab model combined with ab initio calculations \cite{sohier2016two,sohier2017breakdown,sio2022unified}. However, it comes as a surprise that a genuine 2D theory of long wave length LO phonons and electron-phonon interaction in atomically thin polar insulators is still lacking. The present work aims to fill this gap.    
We propose a macroscopic approach to optical phonon modes in 2D polar materials, which generalizes the classical textbook theory of 3D polar crystals \cite{kittel1987quantum}. 
This provides a unified theoretical framework for describing 2D polarons and exciton-polarons (see Fig.~\ref{fig:1} (a)). We derive the electron-phonon Hamiltonian and adopt the intermediate coupling scheme of Lee, Low, and Pines (LLP) \cite{lee1953motion} to obtain the polaron binding energy and effective mass.
We then analyze the exciton problem, where, similarly to the bulk situation \cite{pollmann1977effective}, we obtain an effective electron-hole interaction potential that includes polaronic corrections.
We find that polaron effects substantially modify the exciton binding energy and the shape of wave function, which has implications for interpreting the optical spectra of 2D polar semiconductors.

\textit{Phonon dispersion}.
We consider 2D optical phonon modes in a polar monolayer located at the plane $z=0$.
The Lagrangian density of the system in terms of the mass-weighted displacement field $\bm{\xi}$ reads 
\begin{equation}
    \label{eq:Lagrangian}
    \mathcal{L}= \left[\frac{1}{2}\dot{\bm{\xi}}^2 - W^{\rm 2D}\right]\delta(z) 
    + \varepsilon_0\varepsilon \frac{\mathbf{E}^2(\mathbf{r}, z)}{2},
\end{equation}
where $W^{\rm 2D}$ is the 2D potential energy density of the ionic system, $\varepsilon$ is the dielectric constant of the environment, 
$\varepsilon_0$ is the vacuum permittivity, 
$\mathbf{E}=-\nabla\varphi$ is the electric field, and the last term describes the contribution of electromagnetic field in the bulk environment. Similarly to the 3D case \cite{kittel1987quantum}, the potential energy within the 2D layer has a form fixed by the symmetry \footnote{For simplicity we assume a 2D isotropy of the layer, but this assumption can be readily relaxed.}:
\begin{equation}
    W^{\rm 2D} = 
    \frac{\omega_{\rm t}^2}{2} \bm{\xi}^2 
    - \gamma \bm{\xi} \cdot \mathbf{E}(\mathbf{r},z=0) 
    - \frac{\alpha_\infty}{2}\mathbf{E}^2(\mathbf{r},z=0),
\end{equation}
where the coefficients are uniquely determined from the high and low frequency asymptotic values of the macroscopic 2D polarization $\mathbf{P}=-\partial W^{\rm 2D}/\partial\mathbf{E}$. Specifically, here $\omega_{\rm t}$ is the transverse optical (TO) phonon frequency, $\alpha_\infty$ is the high-frequency 2D polarizability, and $\gamma  = \omega_{\rm t} \sqrt{\alpha_0 - \alpha_\infty}$ where $\alpha_0$ is the static polarizability, see {\color{blue} Supplemental Material (SM) Sec. I} for details.
Apparently, the difference $\alpha_0 - \alpha_\infty$ characterizes the ionic contribution to the polarizability of the monolayer.

The Euler-Lagrange equations, generated by  Eq.~\eqref{eq:Lagrangian}, consist of the equation of motion (EOM) for the 2D displacement $\bm{\xi}(\mathbf{r})$ and the 3D Poisson equation for the electric potential $\varphi(\mathbf{r},z)$. By considering plane-wave solutions, $\bm{\xi}(\mathbf{r}) \sim e^{i \mathbf{k}\cdot\mathbf{r}}$, $\varphi (\mathbf{r},z)\sim e^{i (\mathbf{k}\cdot\mathbf{r} +k_z z)}$,  and defining the in-plane electric potential in the $\mathbf{k}$-space as $\varphi_{\mathbf{k}}^{\rm 2D} = \int \varphi_{ \mathbf{k}, k_z}{\rm d} k_z / (2\pi)$, we find
\begin{align} \label{phi-2D}
\varphi_{\mathbf{k}}^{\rm 2D}   
= -\frac{i\gamma }{4\pi\varepsilon_0 k}\cdot\frac{\mathbf{k} \cdot \bm{\xi}_{\mathbf{k}} }{\varepsilon +  \alpha_\infty k / (2\varepsilon_0)}.
\end{align}
Substitution of this expression into EOM for the displacement, yields the dispersion relation for LO phonons,
\begin{equation}
\label{eq:dispersion}
\omega_{{\rm l},k}^2 = \omega_{\rm t}^2  { \frac{\varepsilon + r_0 k}{\varepsilon + r_\infty k} },
\end{equation}
where screening lengths are defined as
$r_\infty = \alpha_\infty /(2 \varepsilon_0)$, 
$r_0 = \alpha_0 /(2 \varepsilon_0)$.
Equation~\eqref{eq:dispersion} perfectly agrees the dispersion obtained from heuristic arguments supported by direct first principal calculations in Ref.~\cite{sohier2017breakdown} and recently observed experimentally \cite{Li2024experiment}. Remarkably, in our approach, as expected physically in the long wave length limit, the relation between frequencies of TO an LO phonons involves only experimentally accessible macroscopic polarizabilities. The LO-TO degeneracy at $k=0$ as well as a nontrivial dispersion of LO mode are the manifestations of the dielectric screening in 2D. 
It is worth noting that in many instances, the quantity $\varepsilon + \alpha k/2\varepsilon_0$ enters as an effective 2D dielectric constant, and therefore Eq.~\eqref{eq:dispersion} can be viewed as a 2D form of the Lyddane-Sachs-Teller relation.
In Fig.~\ref{fig:1}(b) we visualize the LO phonon dispersion for hBN monolayer \footnote{
For the sake of simplicity in all explicit calculations we assume freestanding atomic monolayers, i.e. $\varepsilon=1$. The remaining parameters are summarized in Table~\ref{table:1}.}.

\begin{table*}[t]
\caption{Calculated polaron binding energies $E^{\rm b}_e$, $E^{\rm b}_h$,
statically screened ($E_{\rm X}^{0}$), exciton-polaron ($E_{\rm X}^{\rm b}$), and bare ($E_{\rm X}^{\infty}$)
exciton binding energies,
renormalized effective masses $m_e^* = m_e (1+\chi_e)$, $m_h^* =m_h (1+\chi_h)$.
Material parameters taken from experimental and first-principal calculations data for a set of 2D polar materials: 
TO phonon energy $\hbar \omega_{\rm t}$, 
screening lengths $r_0$, $r_\infty$\footnote{the screening lengths are extracted from Ref. \cite{sio2023polarons} as
$r_{0[\infty]} = \epsilon_{0[\infty]} d/2$, 
where $\epsilon_{0[\infty]}$ is the static [high-frequency] dielectric constant, $d$ is the monolayer thickness.}, 
electron and hole effective masses $m_e$, $m_h$ (taken from Ref. \cite{sio2023polarons}).
} 
\label{table:1}
\begin{ruledtabular}
\begin{tabular}{lcclccccc}
Material & $E^{\rm b}_e$ (meV)   & $E^{\rm b}_h$ (meV) & $E_{\rm X}^{0}/E_{\rm X}^{\rm b}/E_{\rm X}^{\infty}$ (meV) & $m_e/m_e^*$ &$m_h/m_h^*$ & $\hbar\omega_{\rm t}$ (meV)  & $r_0$ (nm)    &$r_\infty$ (nm)       \\[3pt]\hline
h-BN & $129.3$   & $116.3$ & $1617/1865/2007$ &$0.83$$/1.00$ &$0.65$$/0.77$ & $172.3$\cite{sohier2017breakdown}            &$1.076$ \cite{sohier2017breakdown}        &$0.780$ \cite{sohier2017breakdown}        \\
GaN  & $55.16$ &$132.0$  & $1304/1586/1660$ &$0.24$$/0.29$ &$1.35$$/2.02$ &  $73.30$\cite{Sanders2017}  &$1.109$        &$0.756$    \\
AlN  & $105.4$ & $186.5$ & $2064/2873/2839$  &$0.51$$/0.71$ &$1.49$$/2.53$ & $74.15$\cite{doi:10.1021/acs.jpcc.6b09706}   &$0.763$        &$0.466$  \\
HfSe$_2$ & $89.56$ & $98.51$ & $127.5/306.9/421.4$ &$0.18$$/0.43$ &$0.23$$/0.57$ & $11.10$\cite{Li_2024} &$21.75$ &$4.246$   \\
HfS$_2$ & $123.2$ & $155.1$ & $197.8/434.2/615.6$ &$0.24$$/0.54$ &$0.44$$/1.07$ & $17.81$\cite{doi:10.1021/acsomega.1c04286}   &$13.98$        &$2.974$  \\
ZrS$_2$ & $142.9$& $133.7$  & $238.5/420.6/619.2$ &$0.31$$/0.58$ &$0.26$$/0.48$  & $29.82$\cite{Pandit_2021} &$10.62$ &$2.826$  \\
\end{tabular}
\end{ruledtabular}
\end{table*}
\begin{figure}[t]
\includegraphics[width=1\columnwidth]{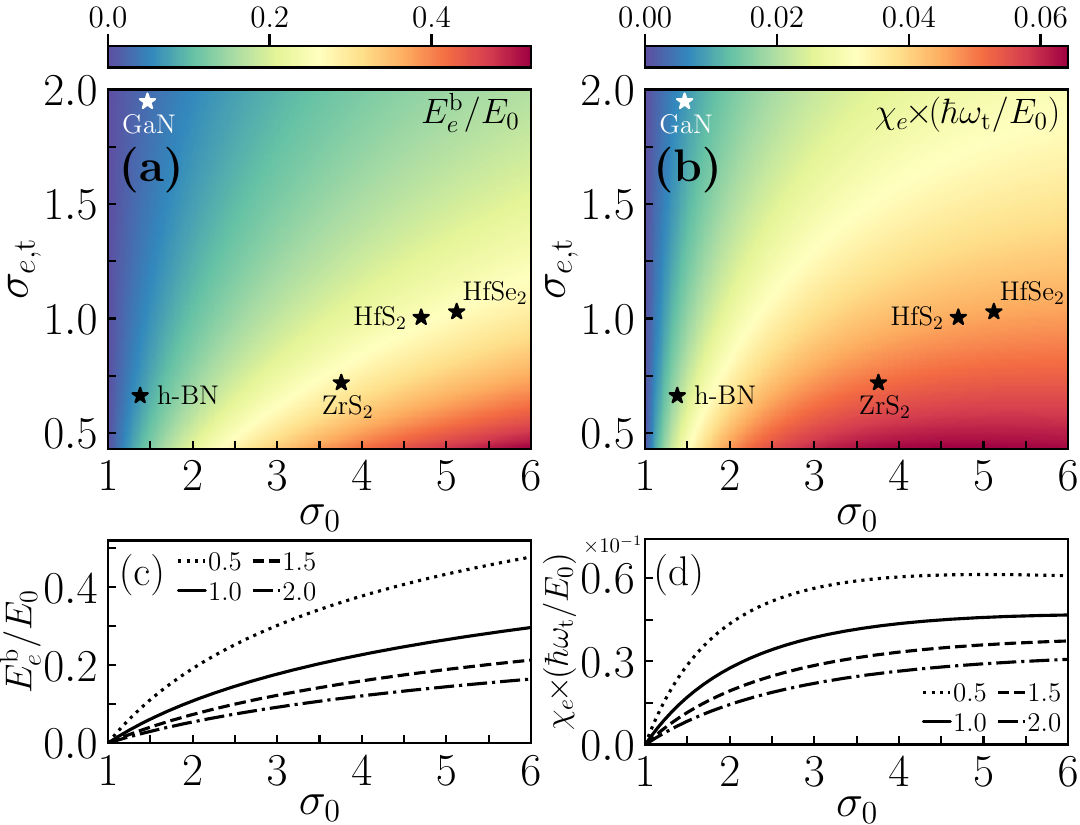}
\caption{(a) The dimensionless binding energy, and (b) mass renormalization rate versus the 
screening length ratios $\sigma_0$, $\sigma_{e, {\rm t}}$.
(c) The dimensionless binding energy, and (d) mass renormalization rate versus $\sigma_0$ for fixed $\sigma_{e, {\rm t}}$, as shown in legend.
Asterisks in panels (a), (b) illustrate the representative polar atomic monolayers.
}
\label{fig:2}
\end{figure}

\textit{Fr\"ohlich Hamiltonian and polaron problem}.
The electric potential of Eq.~\eqref{phi-2D}, produced by the ionic displacement $\bm{\xi}$, acts on the electrons, generating electron-phonon coupling. Formally, the electron-phonon Hamiltonian is derived by inserting the functional $\varphi[\bm{\xi}_\mathbf{k}]$ into the Lagrangian \eqref{eq:Lagrangian} and canonically quantizing the displacement field, see {\color{blue} SM, Sec. II, III, IV A}.
We stress that the last term in Eq.~\eqref{eq:Lagrangian} is essential to restore the correct Hamiltonian of LO phonons.
The final electron-phonon Hamiltonian reads
\begin{align}
    \label{eq:H_e-ph}
    \hat{H} = \frac{\hat{\mathbf{p}}_e^2}{2m_e} + \sum_{\mathbf{k}} \left( \hbar \omega_{{\rm l},k}\hat{a}^\dagger_{\mathbf{k}}\hat{a}_{\mathbf{k}}  
    +\left(V_k e^{i\mathbf{k}\cdot\mathbf{r}_e}\hat{a}_{\mathbf{k}} 
+ {\rm h.c.} \right) \right) 
\end{align}    
where $m_e$, $\hat{\mathbf{p}}_e$, $\mathbf{r}_e$ are the electron mass, momentum, and position, respectively, and $\hat{a}_{\mathbf{k}}$ annihilates a LO phonon with momentum $\mathbf{k}$. 
The electron-phonon interaction is given by $-e \varphi^{\rm 2D}(\mathbf{r}_e)$, leading to
the coupling amplitude,
\begin{align} \label{e-ph-coupling}
    V_k = i \omega_{\rm t} 
    \sqrt{ \frac{\hbar e^2(r_0 - r_\infty) }{4 A  \varepsilon_0 \omega_{{\rm l},k} } } \frac{ 1}{\varepsilon +  r_\infty k },  
\end{align}
which depends on the LO phonon dispersion and contains a typical 2D screening factor. 
Here $A$ denotes the normalization area.

The Hamiltonian of Eq.~\eqref{eq:H_e-ph} contains four length scales: two screening lengths $r_0$ and $r_\infty$, the oscillator scale 
$r_{i, {\rm t}} = \sqrt{\frac{\hbar }{2 \omega_{\rm t} m_i}}$, and the Bohr radius $a_{i,{\rm B}}=\frac{4\pi\varepsilon_0\hbar^2}{e^2m_i}$, where $i = e \, [h]$ for electron [hole].
Their independent ratios, e.~g., $\sigma_0 =  r_0 / r_\infty$, $\sigma_{i, {\rm t}} = \varepsilon r_{i, {\rm t}} / r_\infty$, and $\sigma_{i,{\rm B}}=a_{i,{\rm B}}/r_\infty$ define three independent parameters, characterizing the 2D Fr\"ohlich polaron problem, which should be contrasted to the 3D problem determined by a single dimensionless coupling constant. The reason for this complication is a non-locality of 2D screening. 

We analyze the properties of 2D polaron by adopting the LLP intermediate coupling approach \cite{lee1953motion}, which is expected to be accurate in most realistic situations.
Firstly, we eliminate the electron coordinate by performing a unitary transformation with the operator
$\hat{S} = \exp  \left\{ i \left( \mathbf{p}/h 
-  \sum_{\mathbf{k}} \mathbf{k}\hat{a}^\dagger_{\mathbf{k}} \hat{a}_{\mathbf{k}}  \right) \cdot \mathbf{r}_e \right\}$.
Next, the phonon field is shifted with 
$\hat{U} = \exp\left\{\sum_{\mathbf{k}} \left( f_\mathbf{k}\hat{a}^\dagger_{\mathbf{k}}
- f^*_\mathbf{k}\hat{a}_{\mathbf{k}}\right) \right\}$, 
acting as
$\hat{U}^{-1} \hat{a}_{\mathbf{k}} \hat{U} = \hat{a}_{\mathbf{k}} + f_\mathbf{k}$, where $f_\mathbf{k}$ is a $C$-number serving as a variational parameter. The variational anzats reads $|\psi \rangle = \hat{U}  | \psi_0 \rangle$, assuming $\hat{a}_{\mathbf{k}} |\psi_0 \rangle = 0$.
The variational minimization leads to the following expression for polaron energy (see {\color{blue} SM  Sec. IV B} for the details):
\begin{equation}
    \label{eq:Epol}
    E_e^{\rm pol}  \approx -E_0 I_{\rm b} (\sigma_0, \sigma_{e,{\rm t}}) 
    + \frac{p^2}{2m_e} \frac{1}{1+\chi_e },
\end{equation}
where $ E_0  = \frac{e^2}{4\pi \varepsilon_0 r_\infty}$, 
$\chi_e  =  E_0 I_m (\sigma_0, \sigma_{e,{\rm t}})/\hbar \omega_{\rm t}$, and we defined dimensionless scaling factors
\begin{align}
     I_{\rm b} &= 
     \int\limits_0^\infty     \frac{(\sigma_0-1)x}{2 (1 +  \sigma_0 x ) (1 +  x) \left[1+ \sigma_{e, {\rm t}}^2 x^2  \sqrt{\frac{1 +  x }{1 +  \sigma_0 x }}\right]} \, {\rm d} x, \label{eq:I1}\\[6pt]
     I_m &=
     \int\limits_0^\infty     \frac{ \sigma_{e, {\rm t}}^2 (\sigma_0-1) x^3}{ (1 +  \sigma_0 x )^2 \left[ 1+ \sigma_{e, {\rm t}}^2 x^2  \sqrt{\frac{1 +  x }{1 +  \sigma_0 x }} \right]^3 } \, {\rm d} x. \label{eq:int}
\end{align}
The first term in Eq.~\eqref{eq:Epol} defines the polaron binding energy $E^{\rm b}_e = E_0 I_{\rm b} (\sigma_0, \sigma_{e,{\rm t}})$, and the second term its effective mass $m_e^* = m_e (1+\chi_e)$. Within the LLP scheme, the dependence on the Bohr radius $\sigma_B$ trivially factors as $E_0$ in Eq.~\eqref{eq:Epol}. A nontrivial dependence on the other two parameters is encoded in the scaling factors
$I_{\rm b} (\sigma_0, \sigma_{e,{\rm t}})$ and
$I_m(\sigma_0,\sigma_{e,{\rm t}})$, which, as shown in
Figs.~\ref{fig:2} (a), (b), grow monotonically with an increase in the parameter $\sigma_0-1\sim \alpha_0 -\alpha_\infty$ that characterizes polarizability of the lattice and the coupling strength. 
However, in contrast to 3D the observed growth is highly nonlinear for $\sigma_0-1>1$, as illustrated in
Figs.~\ref{fig:2} (c), (d) for several fixed values of $\sigma_{e,{\rm t}}$.
The strong deviation from linearity for larger $\sigma_0$ comes from the denominators in Eqs.~\eqref{eq:int}, and is a peculiarity of 2D materials.
This is remarkably different from bulk polarons, where at the LLP level, the binding energy and mass renormalization scale linearly with polarizability of the lattice \cite{lee1953motion}.
The length scale $r_{e,t} \propto 1/\sqrt{\omega_{\rm t} m_e }\propto \sigma_{e, {\rm t}}$ roughly determines the spatial extension of the polaron. This explains the reduction of polaronic effects with the increase of $\sigma_{e, {\rm t}}$, see Fig.~\ref{fig:2}. 

The binding energies and effective masses of polarons, computed from Eqs.~\eqref{eq:Epol}-\eqref{eq:int} for several representative materials, are shown in Table~\ref{table:1}. Unlike the case of 3D polarons \cite{lee1953motion}, in general, the final results involve integrals that are not expressible in elementary functions. However, for realistic not too large values of $\sigma_{e, {\rm t}}$, the following substitution $\sqrt{(1+x)/(1+\sigma_0x)} \approx 1/\sqrt{\sigma_0}$, in the integrand in Eq.~\eqref{eq:I1} can be used to get a simplified form,
\begin{multline}
    I_{\rm b} \approx  \frac{\sigma_0-1}{2(\sigma_0^2+\Tilde{\sigma}_{e, {\rm t}}^2)} 
    \left[ \frac{\pi  (\sigma_0+1) \Tilde{\sigma}_{e, {\rm t}} }{2 \left(\Tilde{\sigma}_{e, {\rm t}}^2+1\right)} \right. \\
    \left. +\frac{\left(\Tilde{\sigma}_{e, {\rm t}}^2-\sigma_0\right) \log \Tilde{\sigma}_{e, {\rm t}} }{\Tilde{\sigma}_{e, {\rm t}}^2+1}
    -\frac{\sigma_0 \log \sigma_0}{\sigma_0-1} \right] ,    
\end{multline}
where $\Tilde{\sigma}_{e, {\rm t}} = \sigma_{e, {\rm t}} / \sigma_0^{1/4}$.
This expression represents an excellent estimate for the polaron binding energy, with a few percent relative error in most realistic situations.
Similarly, an approximate analytic form of the mass scaling factor $I_m$ can be derived; see {\color{blue} SM, Sec.~IV~B}.

The polaron energies in Table~\ref{table:1} are notably different from those obtained in Ref.~\cite{sio2023polarons} from a generalized Landau-Pekar model. In principle, this is not surprising because the adiabatic Landau-Pekar approximation is known to underestimate the binding energy for weak and moderate coupling strengths \cite{kittel1987quantum,devreese2016fr}. In particular, our result for the binding energy in h-BN is almost an order of magnitude larger than that in Ref.~\cite{sio2023polarons}. Admittedly, both LLP and Landau-Pekar approximations are variational and therefore both provide upper bounds to the exact energy. In addition, even within the formal Landau-Pekar approach, our Hamiltonian Eq.~\eqref{eq:H_e-ph} produces results different from the model in \cite{sio2023polarons}. The reason is the presence of the full phonon dispersion in Eq.~\eqref{e-ph-coupling}, missing in \cite{sio2022unified,sio2023polarons}, but becoming important for a moderately strong coupling (see {\color{blue} SM, Sec.~IV~C}).

\begin{figure}[t]
\includegraphics[width=1\columnwidth]{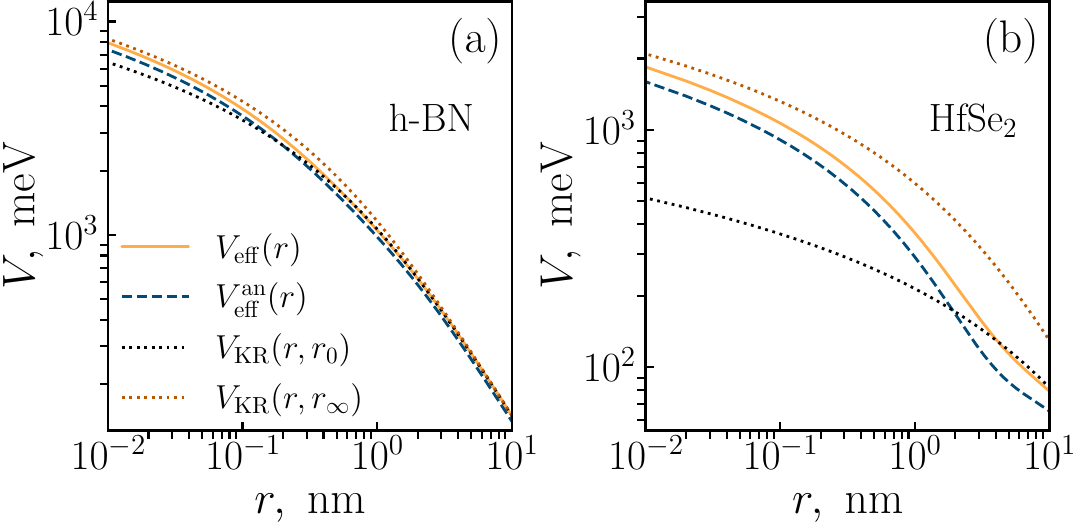}
\caption{Effective potential of electron-hole interaction $V_{\rm eff}$ (solid curve) and its approximate form $V_{\rm eff}^{\rm an}$ (dashed curve) for  (a) h-BN, and (b) HfSe$_2$ monolayers. 
The dotted curves show the KR potentials $V_{\rm KR}(r,r_\infty)$ and $V_{\rm KR}(r,r_0)$. }
\label{fig:3}
\end{figure}
\begin{figure}[b]
\includegraphics[width=1\columnwidth]{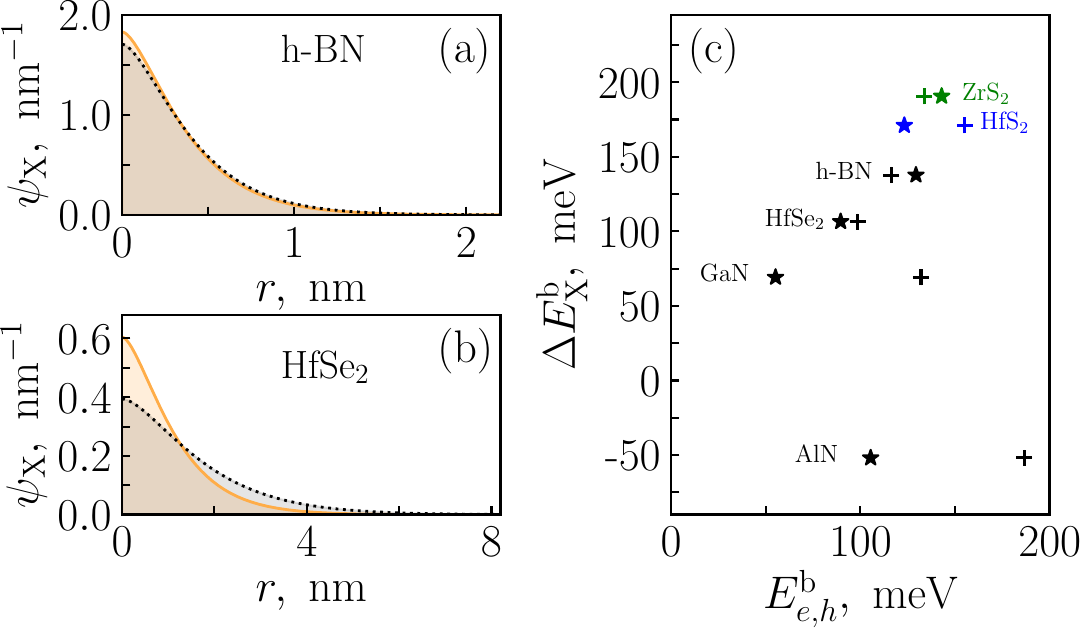}
\caption{
Map of 2D materials illustrating the relation of polaron and exciton-polaron binding energies.
Here $\Delta E^{\rm b}_{\rm X} =  E^{\rm \infty}_{\rm X} -  E^{\rm b}_{\rm X}$. "*" and "+" stand for electron and hole polaron binding energies, respectively.
}
\label{fig:4}
\end{figure}

\textit{Exciton-polaron}. 
When an exciton forms in a 2D polar crystal, both electron and hole couple to LO phonons. 
The resulting composite quasiparticle, an exciton dressed by a phonon cloud, is referred to as an exciton-polaron. 
The Hamiltonian of an electron-hole pair interacting with LO phonons is
\begin{align}
\hat{H}_{\rm X} &= \frac{\hat{\mathbf{p}}_e^2}{2m_e}
+\frac{\hat{\mathbf{p}}_h^2}{2m_h}
+V_{\rm KR}( |\mathbf{r}_e-\mathbf{r}_h|,r_\infty) 
+ \sum_{\mathbf{k}}\hbar\omega_{{\rm l},k}\hat{a}^\dagger_{\mathbf{k}}\hat{a}_{\mathbf{k}} \notag \\
&+ \sum_{\mathbf{k}}\left( 
V_k \hat{a}_{\mathbf{k}} \left(e^{i\mathbf{k}\cdot\mathbf{r}_e} - e^{i\mathbf{k}\cdot\mathbf{r}_h} \right) + {\rm h.c.} \right) ,
\end{align}
where $V_{\rm KR} $ is the "bare" 2D Coulomb (Keldysh-Rytova) potential \cite{Keldysh1979,Rytova1967,cudazzo2011dielectric} with a purely electronic screening,
\begin{equation}    
\!V_{\rm KR} (r,r_\infty)\! =\! -\frac{e^2}{4\pi\varepsilon_0} \frac{\pi}{2 r_\infty}\! \left[ H_0\!\left(\! \frac{\varepsilon r}{r_\infty}\! \right) - Y_0\!\left(\! \frac{\varepsilon r}{r_\infty}\! \right) \right],
\end{equation}
where $H_0$, $Y_0$ are the Struve function, and the Bessel function of the second kind. We then follow a variational approach developed for 3D exciton-polarons \cite{pollmann1977effective,kane1978pollmann}, developed as a generalization of the single polaron LLP theory \cite{lee1953motion}. The main feature of excitonic generalization is that the variational variable $F_\mathbf{k}(\mathbf{r})$, introduced through a shift transformation $\hat{U}^{-1} \hat{a}_{\mathbf{k}} \hat{U} = \hat{a}_{\mathbf{k}} + F_\mathbf{k}(\mathbf{r})$, becomes a function of the relative coordinate $\mathbf{r}=\mathbf{r}_e-\mathbf{r}_h$. The variational energy, which also depends on the exciton wave function $\psi_X(\mathbf{r})$, takes the form (see \cite{kane1978pollmann} and {\color{blue} SM}):
%
\begin{align}
    E_{\rm X} &= \left\langle \psi_{\rm X} \left|
    \hat{\mathbf{p}}^2 /2\mu
    +V_{\rm KR} (r,r_\infty)  
    +W_{\rm X} (\mathbf{r})
     \right|\psi_{\rm X} \right\rangle , 
\end{align}
where $\mu = m_e m_h /M$, $M= m_e + m_h$ are the exciton reduced and total mass, and
\begin{multline}
    W_{\rm X}(\mathbf{r})
    =\sum_k \Big\{\frac{\hbar^2 }{2\mu}|\nabla F_{\mathbf{k}}|^2 + \left( \hbar \omega_{{\rm l}, k}  + \frac{\hbar^2 k^2}{2M}
    \right) |F_\mathbf{k}|^2 \\
    -\left[ V_k \left(e^{i\frac{m_e}{M}\mathbf{k}\cdot\mathbf{r}} - e^{i\frac{m_h}{M}\mathbf{k}\cdot\mathbf{r}} \right) F_{\mathbf{k}} +{\rm c.c.} \right]\Big\} , 
\end{multline} 
Strictly speaking, the presence of the  relative momentum operator $\hat{\mathbf{p}}$ couples the displacement field $F_{\mathbf{k}}(\mathbf{r})$ and the exciton wave function $\psi_{\rm X}$. For simplicity, we ignore this mixing, and apply the variational minimization as 
$\delta W_{\rm X}/ \delta F_{\mathbf{k}} =0$, which determines an effective electron-hole interaction potential $V_{\rm eff}(\mathbf{r})$ dressed by LO phonons. As a result, the exciton energy
is computed as follows,
\begin{align}
    E_{\rm X} &= -E^{\rm b}_e - E^{\rm b}_h + \left\langle \psi_{\rm X} \left|
    \hat{\mathbf{p}}^2/2\mu^*
    +V_{\rm eff} (r)  
    \right|\psi_{\rm X} \right\rangle,
    \label{eq:EXb}
\end{align}
where the negative of the last term defines the binding energy $E_{\rm X}^{\rm b}$ of polaron-exciton, and $\mu^* = m_e^* m_h^* / (m_e^*+m_h^*)$ is the renormalized mass \footnote{Such renormalization of the reduced mass naturally emerges in a more accurate treatment of the variational problem \cite{kane1978pollmann}}.
The exact form of $V_{\rm eff} (r)$ can be found in {\color{blue} SM, Sec.~V}. In the practically relevant case of $\Tilde{\sigma}_{e, {\rm t}}^2, \Tilde{\sigma}_{h, {\rm t}}^2 < 1$, it simplifies as follows,
\begin{multline}
V_{\rm eff}^{\rm ap}(r) \approx V_{\rm KR}(r,r_0) \\
+ E_0 \left[\frac{m_h}{\Delta m}\Phi_{\tilde{\sigma}_{h,t}}\!\! \left(\frac{\varepsilon r}{r_\infty} \right)
-\frac{m_e}{\Delta m} \Phi_{\tilde{\sigma}_{e,t}}\!\! \left(\frac{\varepsilon r}{r_\infty} \right)\right],
\label{eq:Veff}
\end{multline}
where $\Delta m=m_h-m_e$, and the function
\begin{align}
    \Phi_{\nu}( x ) =
    \frac{ \sigma_0 -1}{2(1+\nu^2) (\sigma_0^2+\nu^2)} 
    \big\{ 2 (\nu^2-\sigma_0) K_0\left(x/\nu\right) \notag \\
     +\pi \nu (1+\sigma_0) 
    \left[ I_0\left( x/\nu\right)
    -L_0\left( x/\nu\right) \right] 
      \big\}
\end{align}
encodes phonon-induced corrections.
Here $I_0$, $K_0$ are the modified Bessel functions of the first and the second kind, respectively, and $L_0$ is the modified Struve function. Equation~\eqref{eq:Veff} represents a 2D counterpart of the Pollmann-Büttner potential, characterizing exciton-polarons in 3D \cite{pollmann1977effective,baranowski2020excitons}.

The potential $V_{\rm eff}(r)$ for hBN and HfSe$_2$ monolayers, is shown in Fig.~\ref{fig:3} (a) and (b), respectively. Apparently, the polaron effects are more pronounced in HfSe$_2$ due to stronger coupling with $\sigma_0 \gg 1$. A generic feature of $V_{\rm eff}(r)$ is that it interpolates between the "bare", electron-only screened KR potential, $V_{\rm eff}(r) \rightarrow V_{\rm KR} (r,r_\infty)$, near the origin, and the fully screened potential, $V_{\rm eff}(r) \rightarrow V_{\rm KR} (r,r_0)$, at large distances. This translates into exciton properties, defined by Eq.~\eqref{eq:EXb}.
In particular, in the strongly coupled case of HfSe$_2$ the exciton-polaron binding energy is strongly reduced to 306.9~meV from $E_{\rm X}^\infty=421.4$~meV for the bare exciton; see Table~\ref{table:1}. However, this value is still significantly higher than $E_{\rm X}^0=127.5$~meV obtained naively using the statically screened KR potential $V_{\rm KR} (r,r_0)$.  
In hBN monolayer the polaron screening effects are less pronounced, but the reduction of the binding energy is still sizable, 1865~meV vs 2007~meV for the bare exciton. The data for polaron and exciton-polaron binding in a range of materials are summarized in Fig.~\ref{fig:4}.


\textit{Conclusion}. 
In conclusion, we developed a macroscopic theory of LO phonons, polarons and exciton-polarons in 2D polar materials. Our simple analytic consideration, involving only intrinsic 2D polarizabilities, perfectly agrees with the results of first-principal calculations and recent experiments \cite{sohier2017breakdown,Li2024experiment}.
The variational treatment of the polaron problem results in a semi-analytical expression that defines an upper bound for the energy of Fr{\"o}hlich polaron, which is strongly different from its bulk counterpart \cite{lee1953motion}. The extension of the variational approach to the exciton-polaron problem reveals a modification of the electron-hole interaction by LO phonons, which leads to a dramatic change of the exciton binding energy. Our general framework opens a way for the application of more advanced mathematical techniques to study polaron effects in 2D materials. Apparently, the present purely 2D theory can be readily extended to account for the layered structure of hybrid perovskites \cite{blancon2018scaling}.

\textit{Acknowledgments}. The research is supported by the Ministry of Science and Higher Education of the Russian Federation (Goszadaniye) Project No. FSMG-2023-0011.
V.S. acknowledges the support of “Basis” Foundation (Project No. 22-1-3-43-1). 
The work of A.K. is supported by the Icelandic Research Fund (Ranns\'oknasj\'o{\dh}ur, Grant No.~2410550). IVT acknowledges support from the Spanish MCIN/AEI/
10.13039/501100011033 through the project PID2023-148225NB-C32, and the Basque Government (Grant No. IT1453-22).

\clearpage

\onecolumngrid

\renewcommand{\thefigure}{S\arabic{figure}}
\renewcommand{\theequation}{S\arabic{equation}}
\setcounter{figure}{0}
\setcounter{equation}{0}

{ \bf  \centering SUPPLEMENTAL MATERIAL for: \\ 
\begin{center} Polarons and Exciton-Polarons in Two-Dimensional Polar Materials \end{center} }

{   
\begin{center} V.~Shahnazaryan$^1$,
A.~Kudlis$^{2,1}$, and
I.~Tokatly$^{3,4,5}$
\end{center}
}

\begin{center} $^1$\textit{Abrikosov Center for Theoretical Physics,  Dolgoprudnyi, Moscow Region 141701, Russia} \end{center}

\vspace{-15 pt}

\begin{center} $^2$\textit{Science Institute, University of Iceland, Dunhagi 3, IS-107, Reykjavik, Iceland} \end{center}

\vspace{-15 pt}

\begin{center} $^3$\textit{Departamento de Polímeros y Materiales Avanzados: Física,} \end{center}

\vspace{-15 pt}

\begin{center} \textit{Química y Tecnología, Universidad del País Vasco, } \end{center}

\vspace{-15 pt}

\begin{center} \textit{Avenida Tolosa 72, E-20018 San Sebastián, Spain} \end{center}

\vspace{-15 pt}

\begin{center} $^4$\textit{Donostia International Physics Center (DIPC), E-20018 Donostia-San Sebastián, Spain} \end{center}

\vspace{-15 pt}

\begin{center} $^5$\textit{IKERBASQUE, Basque Foundation for Science, Plaza Euskadi 5, 48009 Bilbao, Spain} \end{center}

\section{2D LO phonon dispersion}

\subsection{Lagrangian density}

Let us define the mass-weighted displacement field $\bm{\xi}=\sqrt{nM}\mathbf{S}$, where $\mathbf{S}=\mathbf{U}_+ - \mathbf{U}_-$ is the relative displacement of positively and negatively charged ions,
$n= N / A$ is the sheet density of unit cells, 
and $M= m_+ m_-/(m_+ + m_-)$ is the reduced ionic mass.
The Lagrangian density of the system is given by:
\begin{align}
\mathcal{L} = \left( \frac{1}{2} \dot{\bm{\xi}}^2 - W^{\rm 2D} \right) \delta (z) 
+ \varepsilon_0 \varepsilon \dfrac{\mathbf{E}^2}{2}.
\end{align}
Here the two terms in brackets correspond to surface density of kinetic and potential energies, respectively.
The last term is the electromagnetic energy density of the surrounding media.
The potential energy density accounts for the energy of ionic displacements, and their interaction with the electric field $\mathbf{E}$:
\begin{align}
W^{\rm 2D} &= \frac{1}{2} \omega_{\rm t}^2 \bm{\xi}^2 - \gamma \bm{\xi} \cdot \mathbf{E} ( \mathbf{r}, z=0) - \frac{\alpha_\infty}{2} \mathbf{E}^2( \mathbf{r}, z=0) .
\end{align}
The 2D polarization field $\mathbf{P}^{\rm 2D}$ then reads:
\begin{align}
\mathbf{P}^{\rm 2D} &= -\dfrac{\partial W^{\rm 2D}}{\partial \mathbf{E}}=\gamma \bm{\xi} + \alpha_\infty \mathbf{E}.
\end{align}
%
Taking the derivative of the potential energy with respect to the displacement field, we obtain:
\begin{align}
\frac{\partial W^{\rm 2D}}{\partial \bm{\xi}} &= \omega_{\rm t}^2 \bm{\xi} - \gamma \mathbf{E}.
\end{align}
In the static case ($\omega = 0$), the system is in equilibrium, so the derivative of the potential energy with respect to $\bm{\xi}$ must vanish:
\begin{align}
\frac{\partial W^{\rm 2D}}{\partial \bm{\xi}} &= 0 \implies \omega_{\rm t}^2 \bm{\xi}_0 - \gamma \mathbf{E} = 0.
\end{align}
Solving with respect to the equilibrium displacement $\bm{\xi}_0$, we find:
\begin{align}
\bm{\xi}_0 &= \frac{\gamma}{\omega_{\rm t}^2} \mathbf{E},
\end{align}
which gives the polarization $\mathbf{P}_0^{\rm 2D}$ in equilibrium
\begin{align}
\mathbf{P}_0^{\rm 2D} &=\gamma \bm{\xi}_0 + \alpha_\infty \mathbf{E} = \left( \alpha_\infty + \frac{\gamma^2}{\omega_{\rm t}^2} \right) \mathbf{E} = \alpha_0 \mathbf{E},
\end{align}
where we have defined the static dielectric polarizability $\alpha_0$ as:
\begin{align}
\alpha_0 &= \alpha_\infty + \frac{\gamma^2}{\omega_{\rm t}^2}.
\end{align}
Expressing the coupling constant $\gamma$ in terms of $\omega_t$, $\alpha_0$, and $\alpha_\infty$, we have:
\begin{align}
\gamma &= \omega_{\rm t} \sqrt{\alpha_0 - \alpha_\infty}.
\end{align}


\subsection{Equations of Motion (EOM)}

The equation of motion for the displacement field $\bm{\xi}$ is derived from the Lagrangian:
\begin{align}
\ddot{\bm{\xi}} &= -\omega_{\rm t}^2 \bm{\xi} + \gamma \mathbf{E}(\mathbf{r}, z=0).
\end{align}
Here, the electric field $\mathbf{E}$ depends on the two-dimensional coordinate vector $\mathbf{r}$, and the condition $z=0$ is due to the two-dimensional nature of the system.
Gauss's law relates the divergence of the electric field to the charge density $\rho$:
\begin{align}
\nabla \cdot \mathbf{E}({\mathbf r}, z) &= \frac{\rho}{\varepsilon \varepsilon_0}  \equiv -\frac{1}{\varepsilon\varepsilon_0} \nabla \cdot \mathbf{P}^{\rm 2D} \delta(z).
\end{align}
%


\subsection{Dispersion relation}

Expressing the electric field in terms of the scalar potential $\varphi$
\begin{align}
\mathbf{E} &= -\nabla \varphi,
\end{align}
and substituting into the equations of motion, we get:
\begin{align}
\label{eq:xit}
\ddot{\bm{\xi}} \delta(z) &= \left( -\omega_{\rm t}^2 \bm{\xi} - \gamma \nabla \varphi |_{z=0} \right) \delta(z),
\end{align}
\begin{align}
\nabla^2 \varphi(\bm{r}, z) &= \frac{1}{\varepsilon \varepsilon_0} \nabla \cdot \left( \gamma \bm{\xi} - \alpha_\infty \nabla \varphi |_{z=0} \right) \delta(z).
\end{align}
We seek for plane-wave solutions for the potential and displacement fields:
\begin{align}
\varphi(\mathbf{r}, z) &\sim e^{i(\mathbf{q}\cdot \mathbf{r} + k_z z)}, \quad \xi(\mathbf{r}) \sim e^{i \mathbf{q}\cdot \mathbf{r} }
\end{align}
where $\mathbf{q}$, $k_z$  are the in-plane and normal-to-plane components of the  wavevector.
Applying the Fourier transform, we find:
\begin{align}
-(q^2 + k_z^2) \varphi_{ \mathbf{q}, k_z} &= \frac{1}{\varepsilon \varepsilon_0} i \mathbf{q} \left( \gamma \bm{\xi}_{\mathbf{q}} / (2\pi) - i \alpha_\infty \mathbf{q} \varphi_{\mathbf{q}}^{\rm 2D} \right),
\end{align} 
where
\begin{align}
    \varphi_{\mathbf{q}}^{\rm 2D} = \int \varphi_{ \mathbf{q}, k_z} \frac{ {\rm d} k_z }{2\pi} ,
\end{align}
\begin{align}
    \bm{\xi}_{\mathbf{q}} = \frac{1}{(2\pi)^2} \int e^{-i \mathbf{q} \mathbf{r} }  \bm{\xi} ( \mathbf{r})  {\rm d^2} \mathbf{r} . 
\end{align}
Solving with respect to the potential $\varphi_{\mathbf{q},k_z}$
\begin{align}\label{eq:four_pot}
\varphi_{\mathbf{q},k_z} &= - \frac{1}{\varepsilon \varepsilon_0} \frac{1 }{q^2 + k_z^2} \left( i \gamma \mathbf{q} \cdot \bm{\xi}_{\mathbf{q}} / (2\pi) + \alpha_\infty q^2 \varphi_{\mathbf{q}}^{\rm 2D} \right),
\end{align}
and plugging in, we reach at:
\begin{align}
\varphi_{\mathbf{q}}^{\rm 2D} &=  -\frac{1}{2q \varepsilon \varepsilon_0} \left( i \gamma \mathbf{q} \cdot \bm{\xi}_{\mathbf{q}} / (2\pi) + \alpha_\infty q^2 \varphi_{\mathbf{q}}^{\rm 2D} \right),
\end{align}
finally yielding in
\begin{align}
\varphi_{\mathbf{q}}^{\rm 2D} &= -\frac{1}{4\pi\varepsilon\varepsilon_0}\frac{ \gamma}{1 +  \alpha_\infty q / (2\varepsilon \varepsilon_0)} i \frac{\mathbf{q} \cdot \bm{\xi}_{\mathbf{q}} }{q}  
= -\frac{1}{4\pi\varepsilon\varepsilon_0}\frac{\gamma }{1 +  \alpha_\infty q / (2\varepsilon \varepsilon_0)}
i {\xi}_q^{\rm l} ,
\end{align}
where we define the longitudinal component of the displacement field:
\begin{align}
\xi_q^{\rm l} &= \frac{\mathbf{q} \cdot \bm{\xi}_{\mathbf{q}}}{q}.
\end{align}
We then substitute back into the equation of motion \eqref{eq:xit}, and perform Fourier transform, reaching at
\begin{align}
\frac{1}{2\pi}\omega_{\rm l}^2 \xi_{q,\omega}^{\rm l} 
&=\frac{1}{2\pi} \omega_{\rm t}^2 \xi_{q,\omega}^{\rm l} + i \gamma q \varphi_{\mathbf{q}}^{\rm 2D}.
\end{align}
We thus find the dispersion relation for the longitudinal optical phonon modes:
\begin{align}
\omega_{{\rm l},q}^2 &= \omega_{\rm t}^2 + \frac{1}{2\varepsilon\varepsilon_0}\frac{\gamma^2 q }{1 +  \alpha_\infty q / (2\varepsilon \varepsilon_0)}.
\end{align}
Substituting the expression for $\gamma$, we get:
\begin{align}
\omega_{{\rm l},q}^2 &= \omega_{\rm t}^2 + \frac{1}{2\varepsilon\varepsilon_0}
\frac{\omega_{\rm t}^2 (\alpha_0 - \alpha_\infty) q }{1 +  \alpha_\infty q / (2\varepsilon \varepsilon_0)}.
\end{align}
Therefore, the frequency of the longitudinal optical phonons is:
\begin{align}
\omega_{{\rm l},q}^2 &= \omega_{\rm t}^2 \frac{1 +\alpha_0 q / (2\varepsilon \varepsilon_0)}{1 +  \alpha_\infty q / (2\varepsilon \varepsilon_0)}
= \omega_{\rm t}^2 \frac{\varepsilon +r_0 q }{\varepsilon +  r_\infty q }.
\end{align}
%


\section{LO phonon Hamiltonian}

We introduce generalized momentum as
\begin{align}
    \bm{\eta} = \frac{ \mathcal{L} }{\partial \dot{\bm{\xi}} } = \dot{\bm{\xi}} \delta(z). 
\end{align}
The Hamiltonian density then reads
\begin{align}
    \mathcal{H} = \dot{\bm{\xi}} \bm{\eta} - \mathcal{L}
    = \left( \frac{ \dot{\bm{\xi}}^2 }{2} + 
    \frac{1}{2} \omega_{\rm t}^2 \bm{\xi}^2 - \gamma \bm{\xi} \cdot \mathbf{E} ( \mathbf{r}, z=0) - \frac{\alpha_\infty}{2} \mathbf{E}^2( \mathbf{r}, z=0) \right) \delta(z)
    - \varepsilon_0 \varepsilon \frac{\mathbf{E}^2}{2} ,
\end{align}
resulting in the classical Hamiltonian 
\begin{align}
    H = \int \mathcal{H} {\rm d}^3 r = 
    \int \left[ \left( \frac{ \dot{\bm{\xi}}^2 }{2} + 
    \frac{1}{2} \omega_{\rm t}^2 \bm{\xi}^2 + \gamma \bm{\xi} \cdot \nabla \varphi ( \mathbf{r}, z=0) - \frac{\alpha_\infty}{2} \nabla (\varphi ( \mathbf{r}, z=0) )^2  \right) \delta(z)
    - \varepsilon_0 \varepsilon \frac{ (\nabla \varphi )^2}{2} \right] {\rm d^2} r {\rm d}z .
\end{align}
We evaluate the electric field as
\begin{align}
    \nabla \varphi ( \mathbf{r}, z=0) &=\nabla \left[ \int e^{i \mathbf{q} \cdot \mathbf{r} } \varphi_{ \mathbf{q}, k_z} {\rm d^2} q {\rm d} k_z  \right]
    = 2\pi i \int  \mathbf{q}  e^{i \mathbf{q} \cdot \mathbf{r} }  \frac{-i}{4\pi\varepsilon_0} \frac{ \gamma}{\varepsilon +  r_\infty q }  \frac{\mathbf{q} \cdot \bm{\xi}_{\mathbf{q}} }{q} {\rm d^2} q
    = \frac{\gamma}{2\varepsilon_0} \int \frac{ q}{\varepsilon +  r_\infty q } e^{i \mathbf{q} \cdot \mathbf{r} }  \bm{\xi}_{\mathbf{q}}^{||}  {\rm d^2} q .
\end{align}
Here in the last relation we introduce a definition $\bm{\xi}_{\mathbf{q}}^{||} = \xi_{\mathbf{q}}^{\rm l} \mathbf{q}/q = (\bm{\xi}_{\mathbf{q}} \cdot \mathbf{q}) \mathbf{q} / q^2 $.
Similarly,
\begin{align}
    \nabla \varphi ( \mathbf{r},z) &=\nabla \left[ \int e^{i ( \mathbf{q} \cdot \mathbf{r} + k_z z) } \varphi_{ \mathbf{q}, k_z} {\rm d^2} q {\rm d} k_z  \right]
    = i \int  (\mathbf{q} + \hat{\mathbf{k}}_z k_z)  e^{i (\mathbf{q} \cdot \mathbf{r} + k_z z) } 
    \varphi_{ \mathbf{q}, k_z} {\rm d} k_z  {\rm d^2} q \notag \\
    & = i \int (\mathbf{q} + \hat{\mathbf{k}}_z k_z)  e^{i (\mathbf{q} \cdot \mathbf{r} + k_z z)} 
    \frac{(-1)}{\varepsilon \varepsilon_0} \frac{1 }{q^2 + k_z^2} \left( i \gamma \mathbf{q} \cdot \frac{\bm{\xi}_{ \mathbf{q}} }{2\pi}
    + \alpha_\infty q^2 \varphi_{\mathbf{q}}^{\rm 2D} \right)
    {\rm d} k_z  {\rm d^2} q \notag \\
    &= \frac{\gamma}{2\pi  \varepsilon_0}
    \int   e^{i (\mathbf{q} \cdot \mathbf{r} + k_z z)} 
    \frac{ (\mathbf{q} + \hat{\mathbf{k}}_z k_z)}{q^2 + k_z^2} \frac{ 1}{\varepsilon +  r_\infty q } \mathbf{q} \cdot \bm{\xi}_{ \mathbf{q}}
    {\rm d} k_z  {\rm d^2} q .
\end{align}
We recall that $\bm{\xi} (\mathbf{r}) = \sqrt{ MN / A } \mathbf{S} (\mathbf{r}) $, and introduce a plane-wave expansion
\begin{align}
    \mathbf{S} (\mathbf{r}) = \frac{1}{ \sqrt{N}} \sum_k e^{i \mathbf{k} \cdot \mathbf{r} } \mathbf{X}_{\mathbf{k}},
\end{align}
and evaluate the terms one by one. Namely,
\begin{align}
    \frac{\omega_{\rm t}^2 }{2} \int \bm{\xi}^2 (\mathbf{r}) \delta(z) {\rm d^2} r {\rm d}z 
    &= \frac{\omega_{\rm t}^2 }{2}  \frac{MN}{ A N}
    \int \sum_k \sum_{k'} e^{i (\mathbf{k}+\mathbf{k}') \cdot \mathbf{r} } \mathbf{X}_{\mathbf{k}} \cdot \mathbf{X}_{\mathbf{k}'} {\rm d^2} r
    = \frac{M\omega_{\rm t}^2 }{2} 
    \sum_k  \mathbf{X}_{\mathbf{k}} \cdot \mathbf{X}_{-\mathbf{k}},
\end{align}
\begin{align}
    \frac{1}{2} \int \dot{\bm{\xi}}^2 (\mathbf{r}) \delta(z) {\rm d^2} r {\rm d}z 
    &= \frac{1}{2}  \frac{M N}{A N}
    \int \sum_k \sum_{k'} e^{i (\mathbf{k}+\mathbf{k}') \cdot \mathbf{r} } \dot{\mathbf{X}}_{\mathbf{k}} \cdot \dot{\mathbf{X}}_{\mathbf{k}'} {\rm d^2} r
    = \frac{M}{2} 
    \sum_k  \dot{\mathbf{X}}_{\mathbf{k}} \cdot \dot{\mathbf{X}}_{-\mathbf{k}}
    = \frac{1}{2M} \sum_k \mathbf{P}_{\mathbf{k}} \cdot \mathbf{P}_{-\mathbf{k}},
\end{align}
where $\mathbf{P}_{\mathbf{k}} = M \dot{\mathbf{X}}_{\mathbf{k}}$.
Noting that $\bm{\xi}_\mathbf{k} =\mathbf{X}_\mathbf{k} \sqrt{A M} / (2\pi)^2$, we get
\begin{align}
    \gamma \int \bm{\xi} (\mathbf{r}) \cdot \nabla \varphi ( \mathbf{r}, z=0)  \delta(z) {\rm d^2} r {\rm d}z 
    & = \gamma \int \bm{\xi} (\mathbf{r}) 
    \cdot \frac{\gamma}{2\varepsilon_0} \int \frac{ q}{\varepsilon +  r_\infty q } e^{i \mathbf{q} \cdot \mathbf{r} }  \bm{\xi}_{\mathbf{q}}^{||}  \delta(z) {\rm d^2} r {\rm d}z {\rm d^2} q \notag \\
    & = \frac{\gamma^2}{2\varepsilon_0} \sqrt{\frac{MN}{A}} \frac{1}{\sqrt{N}} 
    \int \sum_{k} e^{i \mathbf{k} \cdot \mathbf{r} } \frac{ q}{\varepsilon + r_\infty q } e^{i \mathbf{q} \cdot \mathbf{r} } \frac{\sqrt{A M }}{(2\pi)^2} \mathbf{X}_{\mathbf{q}}^{||} \cdot \mathbf{X}_{\mathbf{k}}   {\rm d^2} r{\rm d^2} q \notag \\
    &= M \omega_{\rm t}^2 
     \sum_{k}  \frac{ (r_0 -r_\infty) k}{\varepsilon +  r_\infty k }   \mathbf{X}_{\mathbf{k}}^{||} \cdot \mathbf{X}_{-\mathbf{k}}^{||}
\end{align}
\begin{align}
    \frac{\alpha_\infty}{2} \int  \left[ \nabla \varphi ( \mathbf{r}, z=0) \right]^2 \delta(z) {\rm d^2} r {\rm d}z 
    & = \frac{\alpha_\infty}{2} \int  \frac{\gamma^2}{(2\varepsilon_0)^2}
    \frac{ q}{\varepsilon +  r_\infty q } 
    \frac{ k}{\varepsilon +  r_\infty k }
    e^{i (\mathbf{q} +\mathbf{k}) \cdot \mathbf{r} }  \bm{\xi}_{\mathbf{q}}^{||} \cdot \bm{\xi}_{\mathbf{k}}^{||}  \delta(z) {\rm d^2} r {\rm d}z {\rm d^2} q {\rm d^2} k \notag \\
    & = \frac{\alpha_\infty}{2} \frac{\gamma^2}{(2\varepsilon_0)^2} 
    \int  \frac{ q}{\varepsilon +  r_\infty q }
    \frac{ k}{\varepsilon +  r_\infty k }
    \delta(\mathbf{q} +\mathbf{k}) (2\pi)^2
    \frac{A M }{(2\pi)^4} \mathbf{X}_{\mathbf{q}}^{||} \cdot \mathbf{X}_{\mathbf{k}}^{||}   {\rm d^2} q{\rm d^2} k \notag \\
     & = \frac{M}{2} \omega_{\rm t}^2 
    \sum_k  \frac{ (r_0 - r_\infty) k r_\infty k}{(\varepsilon +  r_\infty k )^2}
     \mathbf{X}_{\mathbf{-k}}^{||} \cdot \mathbf{X}_{\mathbf{k}}^{||}  
\end{align}
\begin{align}
    &\frac{\varepsilon\varepsilon_0}{2} \int (\nabla\varphi)^2 {\rm d}^2 r {\rm d} z \notag \\
    & =\frac{\varepsilon\varepsilon_0}{2} \frac{1}{(2\pi)^2} 
    \frac{\gamma^2}{ \varepsilon_0^2} 
    \int   e^{i (\mathbf{q} \cdot \mathbf{r} + q_z z)}
    e^{i (\mathbf{k} \cdot \mathbf{r} + k_z z)}
    \frac{\mathbf{q} + \hat{\mathbf{k}}_z q_z}{q^2 + q_z^2} \cdot
    \frac{ \mathbf{k} + \hat{\mathbf{k}}_z k_z}{k^2 + k_z^2}
    \frac{ 1}{\varepsilon +  r_\infty q }
    \frac{ 1}{\varepsilon +  r_\infty k }
    (\mathbf{q} \cdot \bm{\xi}_{ \mathbf{q}})
    (\mathbf{k} \cdot \bm{\xi}_{ \mathbf{k}}) 
    {\rm d} q_z  {\rm d^2} q {\rm d} k_z  {\rm d^2} k {\rm d}^2 r {\rm d} z \notag \\
    & =  \frac{2\pi \varepsilon \gamma^2}{ 2 \varepsilon_0} 
    \int \delta(\mathbf{q} + \mathbf{k}) \delta(k_z + q_z)  \frac{ \mathbf{q} \cdot \mathbf{k} + q_z k_z}{(q^2 + q_z^2) (k^2 + k_z^2)}
    \frac{ 1}{\varepsilon + r_\infty q }
    \frac{ 1}{\varepsilon + r_\infty k }
    (\mathbf{q} \cdot \bm{\xi}_{ \mathbf{q}})
    (\mathbf{k} \cdot \bm{\xi}_{ \mathbf{k}}) {\rm d} q_z  {\rm d^2} q {\rm d} k_z  {\rm d^2} k \notag \\
    & = \frac{2\pi \varepsilon \gamma^2}{ 2 \varepsilon_0} 
    \int  \frac{k^2 + k_z^2}{ (k^2 + k_z^2)^2}
    \frac{ 1}{(\varepsilon +  r_\infty k)^2} k^2
    \bm{\xi}_{ \mathbf{-k}}^{||} \cdot \bm{\xi}_{ \mathbf{k}}^{||}  {\rm d} k_z  {\rm d^2} k \notag \\
    & = \frac{M}{2}\omega_{\rm t}^2 \varepsilon
    \sum_k  \frac{ (r_0 - r_\infty) k}{(\varepsilon +  r_\infty k )^2} 
     \mathbf{X}_{ \mathbf{-k}}^{||} \cdot \mathbf{X}_{ \mathbf{k}}^{||}
\end{align}

Collecting all the terms, we reach at
\begin{align}
    H &= \frac{1}{2M} 
    \sum_k  \mathbf{P}_{\mathbf{k}} \cdot \mathbf{P}_{-\mathbf{k}} 
    +\frac{M}{2} \omega_{\rm t}^2 
    \sum_k  \mathbf{X}_{\mathbf{k}} \cdot \mathbf{X}_{-\mathbf{k}} 
    + M\omega_{\rm t}^2 
    \sum_k  \mathbf{X}_{\mathbf{k}}^{||} \cdot \mathbf{X}_{-\mathbf{k}}^{||}
    \left[  \frac{ (r_0 -r_\infty) k}{\varepsilon +  r_\infty k }
    -\frac{1}{2} \frac{ (r_0 - r_\infty) k r_\infty k }{(\varepsilon + r_\infty k )^2}
    - \frac{\varepsilon}{2} \frac{ (r_0 - r_\infty) k}{(\varepsilon +  r_\infty k)^2} 
    \right] \notag \\
    &= \frac{1}{2M} 
    \sum_k  \mathbf{P}_{\mathbf{k}} \cdot \mathbf{P}_{-\mathbf{k}}
    + \frac{M}{2}\omega_{\rm t}^2 
    \sum_k  \mathbf{X}_{\mathbf{k}} \cdot \mathbf{X}_{-\mathbf{k}}
    + M\omega_{\rm t}^2 
    \sum_k  \mathbf{X}_{\mathbf{k}}^{||} \cdot \mathbf{X}_{-\mathbf{k}}^{||}
    \frac{ (r_0 -r_\infty) k}{\varepsilon + r_\infty k }
    \left(1 - \frac{1}{2} \frac{ r_\infty k }{\varepsilon + r_\infty k}
    - \frac{\varepsilon}{2} \frac{ 1 }{\varepsilon + r_\infty k}\right)
     \notag \\
    &= \sum_k \left( 
    \frac{1}{2M} \mathbf{P}_{\mathbf{k}}^{\perp} \cdot \mathbf{P}_{-\mathbf{k}}^{\perp}
    + \frac{M  \omega_{{\rm t}}^2 }{2} 
    \mathbf{X}_{\mathbf{k}}^{\perp} \cdot \mathbf{X}_{-\mathbf{k}}^{\perp} \right)
    + \sum_k \left( 
    \frac{1}{2M} \mathbf{P}_{\mathbf{k}}^{||} \cdot \mathbf{P}_{-\mathbf{k}}^{||}
    + \frac{M \omega_{{\rm l},k}^2 }{2}  \mathbf{X}_{\mathbf{k}}^{||} \cdot \mathbf{X}_{-\mathbf{k}}^{||} \right) 
    = H_{\rm TO} + H_{\rm LO},
\end{align}
where $\mathbf{\xi}_{\mathbf{k}}^{\perp} = \mathbf{\xi}_{\mathbf{k}} - \mathbf{\xi}_{\mathbf{k}}^{||}$ is the transverse component of the displacement.


\section{Electron-phonon coupling }

Introducing the notations
\begin{align}
    \mathbf{X}_{\mathbf{k}}^{||} &= \sqrt{ \frac{\hbar}{2M\omega_{{\rm l}, k} } } \left( a_{\mathbf{k}}  + a_{-\mathbf{k}}^\dag  \right) \mathbf{e}_{\mathbf{k}}, \\
    \mathbf{P}_{\mathbf{k}}^{||} &= \frac{1}{i}\sqrt{ \frac{\hbar M\omega_{{\rm l}, k} }{2 } } \left( a_{\mathbf{k}}  - a_{-\mathbf{k}}^\dag  \right) \mathbf{e}_{\mathbf{k}}, 
\end{align}
We can rewrite the classical Hamiltonian in the form
\begin{align}
    \label{eq:HamClass}
    H_{\rm LO} = \sum_k \frac{\hbar \omega_{{\rm l}, k} }{2} 
    ( a_\mathbf{k} a_\mathbf{k}^\dag  + a_{-\mathbf{k}}^\dag a_{-\mathbf{k}} ) 
\end{align}
The longitudinal displacement then takes a form
\begin{align}
    \mathbf{S}^{||} (\mathbf{r}) &= \frac{1}{ \sqrt{N}} \sum_k e^{i \mathbf{k} \cdot \mathbf{r} } \mathbf{X}_{\mathbf{k}}^{||}
    =  \frac{1}{ \sqrt{N}} \sum_k \sqrt{ \frac{\hbar}{2M\omega_{{\rm l}, k} } } (e^{i \mathbf{k} \cdot \mathbf{r} } a_{\mathbf{k}} + e^{i \mathbf{k} \cdot \mathbf{r} } a_{-\mathbf{k}}^\dag ) \mathbf{e}_{\mathbf{k}} \notag \\
    &=  \frac{1}{ \sqrt{N}} 
    \sum_k \sqrt{ \frac{\hbar}{2M\omega_{{\rm l}, k} } }
    \left(e^{i \mathbf{k} \cdot \mathbf{r} }  a_{\mathbf{k}}
    + e^{ - i \mathbf{k} \cdot \mathbf{r} } a_{\mathbf{k}}^\dag
   \right) \mathbf{e}_{\mathbf{k}},
\end{align}
where in the last relation we used the fact $\mathbf{e}_{-\mathbf{k}} = \mathbf{e}_{\mathbf{k}}$ for the given (longitudinal) polarization of displacement field.
Correspondingly, the mass-weighted displacement is
\begin{align}
    \bm{\xi}^{||} (\mathbf{r}) 
    &=  \sqrt{\frac{MN}{A}}  \mathbf{S}^{||} (\mathbf{r})
   = \sum_k \sqrt{ \frac{\hbar}{2A \omega_{{\rm l}, k} } }
    \left(e^{i \mathbf{k} \cdot \mathbf{r} }  a_{\mathbf{k}}
    + e^{ - i \mathbf{k} \cdot \mathbf{r} } a_{\mathbf{k}}^\dag
   \right) \mathbf{e}_{\mathbf{k}}.
\end{align}
The interaction of charge carrier with the electric field by LO phonon is $ -e \varphi(\mathbf{r}_e,0)$, where the LO phonon potential is readily evaluated as
\begin{align}
    \varphi(\mathbf{r},0) &= \int e^{i\mathbf{q}\cdot\mathbf{r}} \varphi_{\mathbf{q},k_z} {\rm d}^2 q {\rm d} k_z
    = \int e^{i\mathbf{q}\cdot\mathbf{r}}  \frac{(-1)}{\varepsilon \varepsilon_0} \frac{1 }{q^2 + k_z^2} \left( i \gamma \mathbf{q} \cdot \frac{\bm{\xi}_{\mathbf{q}} }{2\pi}  + \alpha_\infty q^2 \varphi_{\mathbf{q}}^{\rm 2D} \right) {\rm d}^2 q {\rm d} k_z \notag \\
    &= -\frac{1}{\varepsilon \varepsilon_0} \int e^{i\mathbf{q}\cdot\mathbf{r}}   \frac{1 }{q^2 + k_z^2} \left( i \gamma \mathbf{q} \cdot \frac{\bm{\xi}_{\mathbf{q}} }{2\pi}  - \alpha_\infty q^2 \frac{1}{4\pi\varepsilon_0}\frac{ \gamma}{\varepsilon +  r_\infty q }
    i \frac{\mathbf{q} \cdot \bm{\xi}_{\mathbf{q}} }{q} \right) {\rm d}^2 q {\rm d} k_z \notag \\
    &= -\frac{i\gamma}{2 \varepsilon_0} \int e^{i\mathbf{q}\cdot\mathbf{r}}   \frac{1 }{q} \frac{ 1}{\varepsilon +  r_\infty q}   \mathbf{q} \cdot \bm{\xi}_{\mathbf{q}} {\rm d}^2 q .
\end{align}
Further,
\begin{align}
     \bm{\xi}_\mathbf{q}^{||} 
     =\frac{\sqrt{A M}}{ (2\pi)^2 }  \mathbf{X}_\mathbf{q}^{||}  
     = \sqrt{ \frac{\hbar A}{2\omega_{{\rm l}, q} } } \frac{1}{(2\pi)^2}
    \left(   a_{\mathbf{q}}  +   a_{-\mathbf{q}}^\dag  \right) \mathbf{e}_{\mathbf{q}},
\end{align}
so that
\begin{align}
    \varphi^{\rm 2D}(\mathbf{r}) \equiv \varphi(\mathbf{r},0)   &= -\frac{i\gamma}{2  \varepsilon_0} \int e^{i\mathbf{q}\cdot\mathbf{r}}   \frac{1 }{q} \frac{ 1}{\varepsilon +  r_\infty q }   \mathbf{q} \cdot \sqrt{ \frac{\hbar A}{2\omega_{{\rm l}, q} } } \frac{1}{(2\pi)^2}
    \left(   a_{\mathbf{q}}  +   a_{-\mathbf{q}}^\dag  \right) \mathbf{e}_{\mathbf{q}} {\rm d}^2 q \notag \\
    &= -i\omega_{\rm t}
    \sum_q \sqrt{ \frac{\hbar (r_0 - r_\infty) }{4 \varepsilon_0 A \omega_{{\rm l}, q} } }  
    \frac{ 1}{\varepsilon +  r_\infty q }  
    \left(   a_{\mathbf{q}} e^{i\mathbf{q}\cdot\mathbf{r}} 
    - a_{\mathbf{q}}^\dag e^{-i\mathbf{q}\cdot\mathbf{r}} \right) .
\end{align}
%


\section{Electron-polarons}

\subsection{Hamiltonian and unitary transformations}

The quantization of Eq. \eqref{eq:HamClass} results in Hamiltonian for LO phonons, reading as 
\begin{align}
    \hat{H}_{\rm ph} = \sum_k \frac{\hbar \omega_{{\rm l}, k} }{2} 
    ( \hat{a}_\mathbf{k} \hat{a}_\mathbf{k}^\dag  + \hat{a}_{-\mathbf{k}}^\dag \hat{a}_{-\mathbf{k}} ) 
    = \sum_k \hbar \omega_{{\rm l}, k} 
    ( \hat{a}_\mathbf{k}^\dag \hat{a}_\mathbf{k} + 1/2 )
\end{align}
The overall electron-LO phonon Hamiltonian is then
\begin{align}
    \hat{H}_{\rm e-ph} = \sum_k \hbar \omega_{{\rm l}, k} 
    \hat{a}_\mathbf{k}^\dag \hat{a}_\mathbf{k} 
    + \sum_k  \left( V_k e^{i\mathbf{k}\cdot\mathbf{r}_e} \hat{a}_{\mathbf{k}}  +  V_k^* e^{-i\mathbf{k}\cdot\mathbf{r}_e} \hat{a}_{\mathbf{k}}^\dag \right) 
    +  \frac{\hat{\mathbf{p}}_e^2}{2m_e}
\end{align}
where 
\begin{align}
    V_k = i e  \omega_{\rm t} 
    \sqrt{ \frac{\hbar (r_0 - r_\infty) }{4 \varepsilon_0 A \omega_{{\rm l},k} } } \frac{ 1}{\varepsilon +  r_\infty k }.  
\end{align}
We further apply a variational approach for the calculation of polaron binding energy and charge carrier mass renormalization. 
In virtue of Ref.~\cite{lee1953motion}, we introduce the total momentum operator
\begin{align}
    \hat{\mathbf{p}} = \sum_k \hbar \mathbf{k} 
    \hat{a}_\mathbf{k}^\dag \hat{a}_\mathbf{k} + \hat{\mathbf{p}}_e .
\end{align}
One has
\begin{align}
    [\hat{\mathbf{p}}, \hat{\mathbf{p}}_e^2] =0, \qquad 
   [\hat{\mathbf{p}}, \hat{a}_\mathbf{q}^\dag \hat{a}_\mathbf{q}] =0 ,  \qquad
   [\hat{\mathbf{p}}, \hat{a}_\mathbf{q} e^{i \mathbf{q} \cdot \mathbf{r}_e} ] = \sum_k \hbar \mathbf{k} 
    [\hat{a}_\mathbf{k}^\dag \hat{a}_\mathbf{k}, \hat{a}_\mathbf{q}] e^{i \mathbf{q} \cdot \mathbf{r}_e} 
    + \hat{a}_\mathbf{q} [\hat{\mathbf{p}}, e^{i \mathbf{q} \cdot \mathbf{r}_e}] 
    = 0 , 
\end{align}
so that 
\begin{align}
    [\hat{\mathbf{p}}, \hat{H}_{\rm e-ph}] =0 , 
\end{align}
ensuring the total momentum conservation.
We then  eliminate electron coordinates performing a unitary transformation with operator
\begin{align}
    \hat{S} = e^{ i (\mathbf{p}/\hbar - \sum_k \mathbf{k} 
    \hat{a}_\mathbf{k}^\dag \hat{a}_\mathbf{k}) \cdot {\mathbf{r}_e}} .
\end{align}
One has
\begin{align}
    \Tilde{\hat{\mathbf{p}}}_e &= 
     \hat{\mathbf{p}}_e  +  \mathbf{p} 
    - \hbar \sum_k \hat{a}_\mathbf{k}^\dag \hat{a}_\mathbf{k} \mathbf{k} ,
\end{align}
so that
\begin{align}
    \Tilde{\hat{\mathbf{p}}}_e^2 &= \hat{S}^{-1} \hat{\mathbf{p}}_e^2 \hat{S} = \hat{S}^{-1} \hat{\mathbf{p}}_e \hat{S} \cdot \hat{S}^{-1} \hat{\mathbf{p}}_e \hat{S} 
    = \left( \hat{\mathbf{p}}_e  +  \mathbf{p} 
    - \hbar \sum_k \hat{a}_\mathbf{k}^\dag \hat{a}_\mathbf{k} \mathbf{k} \right)^2 .
\end{align}
Similarly,
\begin{align}
    \Tilde{\hat{a}}_{\mathbf{k}} &= 
    \hat{a}_{\mathbf{k}} e^{-i \mathbf{k} \cdot \mathbf{r}_e} ,
\end{align}
meaning that
\begin{align}
    \Tilde{\hat{a}}_{\mathbf{k}}^\dag 
    & = \hat{a}_{\mathbf{k}}^\dag e^{i \mathbf{k} \cdot \mathbf{r}_e} , \notag \\
    \Tilde{\hat{a}_{\mathbf{k}}^\dag \hat{a}_{\mathbf{k}} } 
    & = \hat{S}^{-1} \hat{a}_{\mathbf{k}}^\dag \hat{S} \cdot \hat{S}^{-1} \hat{a}_{\mathbf{k}} \hat{S} 
    = \hat{a}_{\mathbf{k}}^\dag \hat{a}_{\mathbf{k}}, \notag \\
    \Tilde{\hat{\mathbf{p}}}& =   \hat{\mathbf{p}}_e  +  \mathbf{p} .
\end{align}
Now the transformed Hamiltonian is written as
\begin{align}
    \Tilde{\hat{H}}_{\rm e-ph} = \sum_k \hbar \omega_{{\rm l}, k} 
    \hat{a}_\mathbf{k}^\dag \hat{a}_\mathbf{k} 
    + \sum_k  \left( V_k \hat{a}_{\mathbf{k}}  +  V_k^* \hat{a}_{\mathbf{k}}^\dag \right) 
    +  \frac{ \left( \mathbf{p} - \sum_k \hbar \mathbf{k} 
    \hat{a}_\mathbf{k}^\dag \hat{a}_\mathbf{k} \right)^2}{2m_e},
\end{align}
where we omit $\hat{\mathbf{p}}_e$ provided that we get rid of electron coordinate, so that
$\hat{\mathbf{p}}_e (\hat{S}^{-1} \psi ) =0$.
We apply a variational anzats in a form $\psi = \hat{U} \psi_0$, where $\psi_0$ corresponds to phonon vacuum state $\hat{a}_{\mathbf{k}} \psi_0 =0$, $\langle \psi_0 | \psi_0 \rangle =1$, and unitary operator 
\begin{align}
    \hat{U} = e^{ \sum_k ( \hat{a}_\mathbf{k}^\dag f(\mathbf{k}) - \hat{a}_\mathbf{k} f^*(\mathbf{k}) ) },
\end{align}
where $f(\mathbf{k})$ corresponds to a mean-field amplitude of phonon field minimizing the polaron energy.
One has
\begin{align}
    \hat{U}^{-1} \hat{a}_{\mathbf{k}}^\dag \hat{U} &= \hat{a}_{\mathbf{k}}^\dag 
    - \sum_q [ (\hat{a}_\mathbf{q}^\dag f(\mathbf{q}) - \hat{a}_\mathbf{q} f^*(\mathbf{q}) ),  \hat{a}_{\mathbf{k}}^\dag] + \ldots 
    = \hat{a}_{\mathbf{k}}^\dag - \sum_q (-1) \delta_{\mathbf{q},\mathbf{k}} f^*(\mathbf{q})
    = \hat{a}_{\mathbf{k}}^\dag + f^*(\mathbf{k}),
\end{align}
and
\begin{align}
    \hat{U}^{-1} \hat{a}_{\mathbf{k}}\hat{U} &= \hat{a}_{\mathbf{k}} + f(\mathbf{k})  . 
\end{align}
We evaluate the Hamiltonian terms one by one as
\begin{align}
    \hat{U}^{-1} \sum_k \hbar \omega_{{\rm l}, k} \hat{a}_\mathbf{k}^\dag \hat{a}_\mathbf{k} \hat{U} 
    &= \sum_k \hbar \omega_{{\rm l}, k} 
    \left( \hat{a}_{\mathbf{k}}^\dag \hat{a}_{\mathbf{k}}
    +f(\mathbf{k}) \hat{a}_{\mathbf{k}}^\dag 
    +f^*(\mathbf{k}) \hat{a}_{\mathbf{k}} 
    +f^*(\mathbf{k}) f(\mathbf{k}) \right) \\
    \hat{U}^{-1} \sum_k  \left( V_k \hat{a}_{\mathbf{k}}  +  V_k^* \hat{a}_{\mathbf{k}}^\dag \right) \hat{U} 
    &=  \sum_k  \left( V_k \hat{a}_{\mathbf{k}} + V_k f(\mathbf{k}) 
    +V_k^* \hat{a}_{\mathbf{k}}^\dag + V_k^* f^*(\mathbf{k}) \right)\\
    \hat{U}^{-1} \frac{\mathbf{p}^2}{2m_e} \hat{U} &= \frac{\mathbf{p}^2}{2m_e} \\
    -\frac{1}{m} \hat{U}^{-1} \mathbf{p} \sum_k \hbar \mathbf{k} 
    \hat{a}_\mathbf{k}^\dag \hat{a}_\mathbf{k} \hat{U}
    &= - \frac{\hbar}{m_e} \mathbf{p} \cdot \mathbf{k} \left( \hat{a}_{\mathbf{k}}^\dag \hat{a}_{\mathbf{k}}
    +f(\mathbf{k}) \hat{a}_{\mathbf{k}}^\dag 
    +f^*(\mathbf{k}) \hat{a}_{\mathbf{k}} 
    +f^*(\mathbf{k}) f(\mathbf{k}) \right) 
\end{align}
and
\begin{align}
    &\hat{U}^{-1} \frac{1}{2m_e} \sum_k \sum_{k'} \hbar \mathbf{k} \cdot \hbar \mathbf{k}' 
    \hat{a}_\mathbf{k}^\dag \hat{a}_\mathbf{k} \hat{a}_{\mathbf{k}'}^\dag \hat{a}_{\mathbf{k}'} \hat{U}
    = \frac{\hbar^2}{2m_e} \sum_k \sum_{k'}  \mathbf{k} \cdot  \mathbf{k}' 
    (\hat{a}_{\mathbf{k}}^\dag + f^*(\mathbf{k})) 
    (\hat{a}_{\mathbf{k}} + f(\mathbf{k}))
    (\hat{a}_{\mathbf{k}'}^\dag + f^*(\mathbf{k}')) 
    (\hat{a}_{\mathbf{k}'} + f(\mathbf{k}')) \notag \\
    &= \frac{\hbar^2}{2m_e} \sum_k \sum_{k'}  \mathbf{k} \cdot  \mathbf{k}'
    \left( \hat{a}_\mathbf{k}^\dag \hat{a}_\mathbf{k} \hat{a}_{\mathbf{k}'}^\dag \hat{a}_{\mathbf{k}'}
    +\hat{a}_\mathbf{k}^\dag \hat{a}_{\mathbf{k}'}^\dag \hat{a}_\mathbf{k} f(\mathbf{k}')
    +\hat{a}_\mathbf{k}^\dag \hat{a}_\mathbf{k} \hat{a}_{\mathbf{k}'} f^*(\mathbf{k}')
    +\hat{a}_\mathbf{k}^\dag \hat{a}_\mathbf{k} |f(\mathbf{k}')|^2
    \right. \notag \\
    & \hspace{100 pt} 
    +\hat{a}_\mathbf{k}^\dag \hat{a}_{\mathbf{k}'}^\dag \hat{a}_{\mathbf{k}'} f(\mathbf{k})
    +\hat{a}_\mathbf{k}^\dag \hat{a}_{\mathbf{k}'}^\dag f(\mathbf{k}') f(\mathbf{k})
    +\hat{a}_\mathbf{k}^\dag \hat{a}_{\mathbf{k}'} f^*(\mathbf{k}') f(\mathbf{k}) 
    +\hat{a}_\mathbf{k}^\dag |f(\mathbf{k}')|^2 f(\mathbf{k}) \notag \\
    & \hspace{100 pt} 
    +\hat{a}_{\mathbf{k}'}^\dag \hat{a}_\mathbf{k}  \hat{a}_{\mathbf{k}'} f^*(\mathbf{k})
    +\hat{a}_{\mathbf{k}'}^\dag \hat{a}_\mathbf{k} f(\mathbf{k}') f^*(\mathbf{k})
    +\hat{a}_\mathbf{k} \hat{a}_{\mathbf{k}'} f^*(\mathbf{k}') f^*(\mathbf{k})
    +\hat{a}_\mathbf{k} |f(\mathbf{k}')|^2 f^*(\mathbf{k}) \notag \\
    & \hspace{100 pt} \left.
    +\hat{a}_{\mathbf{k}'}^\dag \hat{a}_{\mathbf{k}'} |f(\mathbf{k})|^2
    +\hat{a}_{\mathbf{k}'}^\dag f(\mathbf{k}') |f(\mathbf{k})|^2 
    +\hat{a}_{\mathbf{k}'} f^*(\mathbf{k}') |f(\mathbf{k})|^2
    + |f(\mathbf{k})|^2 |f(\mathbf{k}')|^2  \right) \notag \\
    & + \frac{\hbar^2}{2m_e} \sum_k \mathbf{k}^2 
    \left( \hat{a}_\mathbf{k}^\dag f(\mathbf{k})
    +\hat{a}_{\mathbf{k}} f^*(\mathbf{k})
    + |f(\mathbf{k})|^2   \right)
\end{align}
Now the Hamiltonian can be presented in the form ordered by the number of phonon creation / annihilation operators:
\begin{align}
    \hat{H}_{\rm e-ph} = \hat{H}_0 + \hat{H}_1 + \hat{H}_2 + \hat{H}_3 + \hat{H}_4 , 
\end{align}
where
\begin{align}
    \hat{H}_0 &= \frac{\mathbf{p}^2}{2m_e} +  \sum_k  \left( V_k f(\mathbf{k}) 
     + V_k^* f^*(\mathbf{k}) \right) + \frac{\hbar^2}{2m_e} \left( \sum_k |f(\mathbf{k})|^2 \mathbf{k} \right)^2 
     +\sum_k |f(\mathbf{k})|^2 \left( \hbar \omega_{{\rm l}, k} - \frac{\hbar}{m_e} \mathbf{p} \cdot \mathbf{k} + \frac{\hbar^2 k^2}{2m_e} \right) \\
    \hat{H}_1 &= \sum_k \left[ V_k 
    + f^* (\mathbf{k}) \left( \hbar \omega_{{\rm l}, k} - \frac{\hbar}{m_e} \mathbf{p} \cdot \mathbf{k} + \frac{\hbar^2 k^2}{2m_e} 
    +\frac{\hbar^2}{m_e} \mathbf{k} \cdot \left( \sum_{k'} |f(\mathbf{k}')|^2 \mathbf{k}' \right) \right)  \right] \hat{a}_{\mathbf{k}} \notag \\
    &+ \sum_k \left[ V_k^* 
    + f (\mathbf{k}) \left( \hbar \omega_{{\rm l}, k} - \frac{\hbar}{m_e} \mathbf{p} \cdot \mathbf{k} + \frac{\hbar^2 k^2}{2m_e} 
    +\frac{\hbar^2}{m_e} \mathbf{k} \cdot \left( \sum_{k'} |f(\mathbf{k}')|^2 \mathbf{k}' \right) \right)  \right] \hat{a}_{\mathbf{k}}^\dag \\
    \hat{H}_2 &= \sum_k  \left( \hbar \omega_{{\rm l}, k} - \frac{\hbar}{m_e} \mathbf{p} \cdot \mathbf{k} 
    +\frac{\hbar^2}{m_e} \mathbf{k} \cdot \left( \sum_{k'}   \mathbf{k}' 
      |f(\mathbf{k}')|^2 \right) \right) \hat{a}_{\mathbf{k}}^\dag \hat{a}_{\mathbf{k}}     \\ 
    &+\frac{\hbar^2}{2m_e} \sum_k \sum_{k'}  \mathbf{k} \cdot  \mathbf{k}' 
    \left( \hat{a}_\mathbf{k}^\dag \hat{a}_{\mathbf{k}'}^\dag f(\mathbf{k}') f(\mathbf{k})
    +\hat{a}_\mathbf{k} \hat{a}_{\mathbf{k}'} f^*(\mathbf{k}') f^*(\mathbf{k})
    +2\hat{a}_\mathbf{k}^\dag \hat{a}_{\mathbf{k}'} f^*(\mathbf{k}') f(\mathbf{k})  \right)\\
    \hat{H}_3 &= \frac{\hbar^2}{m_e} \sum_k \sum_{k'}  \mathbf{k} \cdot  \mathbf{k}' 
    \left( \hat{a}_\mathbf{k}^\dag \hat{a}_\mathbf{k} \hat{a}_{\mathbf{k}'} f^*(\mathbf{k}') 
    + \hat{a}_{\mathbf{k}'}^\dag \hat{a}_\mathbf{k}^\dag \hat{a}_\mathbf{k} f(\mathbf{k}') \right) \\
    \hat{H}_4 &= \frac{ \left( \sum_k \hbar \mathbf{k} \hat{a}_\mathbf{k}^\dag \hat{a}_\mathbf{k} \right)^2}{2m_e}
\end{align}

\subsection{Polaron energy minimization}

Taking into account that $\hat{a}_{\mathbf{k}} \psi_0 =0$, we calculate the variational energy as
\begin{align}
    E &= \langle \psi | \hat{H}_{\rm e-ph} | \psi \rangle
      = \langle \psi | \hat{H}_0 | \psi \rangle \notag \\
      &= \frac{\mathbf{p}^2}{2m_e} +  \sum_k  \left( V_k f(\mathbf{k}) 
     + V_k^* f^*(\mathbf{k}) \right) + \frac{\hbar^2}{2m_e} \left( \sum_k |f(\mathbf{k})|^2 \mathbf{k} \right)^2 
     +\sum_k |f(\mathbf{k})|^2 \left( \hbar \omega_{{\rm l}, k} - \frac{\hbar}{m_e} \mathbf{p} \cdot \mathbf{k} + \frac{\hbar^2 k^2}{2m_e} \right) ,
\end{align}
which can be minimized setting
\begin{align}
    \frac{\delta E}{\delta f(\mathbf{k})} = \frac{\delta E}{\delta f(\mathbf{k}')} = 0,
\end{align}
resulting in
\begin{align}
    & V_k + f^* (\mathbf{k}) \left( \hbar \omega_{{\rm l}, k} - \frac{\hbar}{m_e} \mathbf{p} \cdot \mathbf{k} + \frac{\hbar^2 k^2}{2m_e} 
    +\frac{\hbar^2}{m_e} \mathbf{k} \cdot \left( \sum_{k'} |f(\mathbf{k}')|^2 \mathbf{k}' \right) \right)  =0, \notag \\
    & V_k^* + f (\mathbf{k}) \left( \hbar \omega_{{\rm l}, k} - \frac{\hbar}{m_e} \mathbf{p} \cdot \mathbf{k} + \frac{\hbar^2 k^2}{2m_e} 
    +\frac{\hbar^2}{m_e} \mathbf{k} \cdot \left( \sum_{k'} |f(\mathbf{k}')|^2 \mathbf{k}' \right) \right) =0 .
\end{align}
We note that the obtained expressions correspond to the sum coefficients in $\hat{H}_1$, meaning the exact cancellation of linear in $\hat{a}_{\mathbf{k}}$ terms from the Hamiltonian.
Noting that the only preferred direction is that of $\mathbf{p}$, one may write
\begin{align}
    \beta \mathbf{p} = \sum_{k} |f(\mathbf{k})|^2  \hbar \mathbf{k},
\end{align}
and obtain an equation for amplitude $f(\mathbf{k})$:
\begin{align}
     f (\mathbf{k}) = -\frac{V_k^*}{ \hbar \omega_{{\rm l}, k} - \frac{\hbar}{m_e} \mathbf{p} \cdot \mathbf{k} (1-\beta) + \frac{\hbar^2 k^2}{2m_e} }.  
\end{align}
Plugging in, we get
\begin{align}
    \label{eq:beta}
    \beta \mathbf{p} = \sum_{k} |f(\mathbf{k})|^2 \hbar \mathbf{k}
    =\sum_{k} \frac{ |V_k|^2 \hbar \mathbf{k} }{ \left( \hbar \omega_{{\rm l}, k} - \frac{\hbar}{m_e} \mathbf{p} \cdot \mathbf{k} (1-\beta) + \frac{\hbar^2 k^2}{2m_e} \right)^2 }.
\end{align}

As for energy, we note that
\begin{align}
    \frac{\hbar^2}{2m_e} \left( \sum_{k} |f(\mathbf{k})|^2 \mathbf{k} \right)^2
    &=\frac{\hbar^2}{m_e} \left( \sum_{k} |f(\mathbf{k})|^2 \mathbf{k} \right)^2
    -\frac{\hbar^2}{2m_e} \left( \sum_{k} |f(\mathbf{k})|^2 \mathbf{k} \right)^2
    = \frac{\hbar}{m_e} \beta \mathbf{p} \cdot \sum_k  |f(\mathbf{k})|^2 \mathbf{k} 
    -\frac{1}{2m_e} (\beta \mathbf{p})^2
\end{align}
resulting in
\begin{align}
     E &= \frac{\mathbf{p}^2}{2m_e} (1-\beta^2)
     - \sum_k  \frac{2 |V_k|^2}{ \hbar \omega_{{\rm l}, k} - \frac{\hbar}{m_e} \mathbf{p} \cdot \mathbf{k} (1-\beta) + \frac{\hbar^2 k^2}{2m_e} }
     + \sum_{k} \frac{\hbar}{m_e} \beta \mathbf{p} \cdot \mathbf{k} \frac{ |V_k|^2}{ \left( \hbar \omega_{{\rm l}, k} - \frac{\hbar}{m_e} \mathbf{p} \cdot \mathbf{k} (1-\beta) + \frac{\hbar^2 k^2}{2m_e} \right)^2 }  \notag \\
     &+\sum_k \frac{ |V_k|^2}{ \left( \hbar \omega_{{\rm l}, k} - \frac{\hbar}{m_e} \mathbf{p} \cdot \mathbf{k} (1-\beta) + \frac{\hbar^2 k^2}{2m_e} \right)^2 }
     \left( \hbar \omega_{{\rm l}, k} - \frac{\hbar}{m_e} \mathbf{p} \cdot \mathbf{k} + \frac{\hbar^2 k^2}{2m_e} \right) \notag \\
     &=  \frac{\mathbf{p}^2}{2m_e} (1-\beta^2)
     - \sum_k  \frac{ |V_k|^2}{ \hbar \omega_{{\rm l}, k} - \frac{\hbar}{m_e} \mathbf{p} \cdot \mathbf{k} (1-\beta) + \frac{\hbar^2 k^2}{2m_e} }.
\end{align}

Next, one has
\begin{align}
    |V_k|^2 
     &= \frac{e^2  \omega_{\rm t} \hbar (r_0 - r_\infty) }{A (4 \varepsilon_0)}
     \frac{1}{ \sqrt{ (\varepsilon +  r_0 k ) (\varepsilon +  r_\infty k )} }  \frac{ 1}{ (\varepsilon +  r_\infty k )}.  
\end{align}
%


Expanding the Eq.~\eqref{eq:beta} up to first order in $\mathbf{k}$, we get
\begin{align}
    \beta \mathbf{p} 
    =\sum_{k} \frac{ |V_k|^2 \hbar \mathbf{k} }{ \left( \hbar \omega_{{\rm l}, k} + \frac{\hbar^2 k^2}{2m_e} \right)^3 }
    \left(  \hbar \omega_{{\rm l}, k} 
    + \frac{\hbar^2 k^2}{2m_e}
    +2 \frac{\hbar}{m_e} \mathbf{p} \cdot \mathbf{k} (1-\beta)   \right),
\end{align}
Plugging in the values of $|V_k|^2$,  $\omega_{{\rm l}, k}$, and proceeding with integration, we reach at
\begin{align}
    \beta p 
    &= \frac{A}{(2\pi)^2} \int\limits_0^\infty \int\limits_0^{2\pi} \frac{e^2  \omega_{\rm t} \hbar (r_0 - r_\infty) }{A (4 \varepsilon_0)}
     \frac{1}{ \sqrt{ (\varepsilon +  r_0 k ) (\varepsilon +  r_\infty k )^3} } 
     \left(  \hbar \omega_{{\rm l}, k} 
    + \frac{\hbar^2 k^2}{2m_e}
    +2 \frac{\hbar}{m_e} p k \cos\varphi (1-\beta)   \right)
     \frac{ \hbar k \cos\varphi k {\rm d} k {\rm d } \varphi }{ \left( \hbar \omega_{{\rm l}, k} + \frac{\hbar^2 k^2}{2m_e} \right)^3 } \notag \\
     &= \frac{e^2  \omega_{\rm t} \hbar (r_0 - r_\infty) }{ (8 \pi \varepsilon_0)} 
     \int\limits_0^\infty  
     \frac{1}{ \sqrt{ (\varepsilon +  r_0 k ) (\varepsilon +  r_\infty k )^3} } 
      \frac{\hbar}{m_e} p k (1-\beta)   
     \frac{ \hbar k  k {\rm d} k }{ \left( \hbar \omega_{{\rm l}, k} + \frac{\hbar^2 k^2}{2m_e} \right)^3 }  \notag \\
     &= \frac{e^2   (r_0 - r_\infty) }{ (8 \pi  \varepsilon_0)} \frac{p(1-\beta)}{m_e \omega_{\rm t}^2}
     \int\limits_0^\infty  
     \frac{1}{(\varepsilon +  r_0 k )^2}
     \frac{ k^3  {\rm d} k }{ \left( 1 + r_{e, {\rm t}}^2 k^2 \sqrt{\frac{\varepsilon +  r_\infty k}{\varepsilon +  r_0 k}}
     \right)^3 } , 
\end{align}

Similarly
\begin{align}
    E &\approx  \frac{p^2}{2m_e} (1-\beta^2)
     - \sum_k  \frac{ |V_k|^2}{ \hbar \omega_{{\rm l}, k}
     + \frac{\hbar^2 k^2}{2m_e} }
     \left( 1 +  \frac{ \frac{\hbar}{m_e} \mathbf{p} \cdot \mathbf{k} (1-\beta)}{ \hbar \omega_{{\rm l}, k}
     + \frac{\hbar^2 k^2}{2m_e} }
     + \frac{ \frac{\hbar^2 }{m_e^2} (1-\beta)^2 (\mathbf{p} \cdot \mathbf{k})^2 }{ \left( \hbar \omega_{{\rm l}, k}
     + \frac{\hbar^2 k^2}{2m_e} \right)^2 }
      \right) \notag \\
      &=  \frac{p^2}{2m_e} (1-\beta^2)
     - \frac{A}{(2\pi)^2} \int\limits_0^\infty \int\limits_0^{2\pi} 
     \frac{ |V_k|^2}{ \hbar \omega_{{\rm l}, k}
     + \frac{\hbar^2 k^2}{2m_e} }
     \left( 1 +  \frac{ \frac{\hbar}{m_e} p k \cos\varphi (1-\beta)}{ \hbar \omega_{{\rm l}, k}
     + \frac{\hbar^2 k^2}{2m_e} }
     + \frac{ \frac{\hbar^2 }{m_e^2} (1-\beta)^2 ( p k \cos\varphi)^2 }{ \left( \hbar \omega_{{\rm l}, k}
     + \frac{\hbar^2 k^2}{2m_e} \right)^2 }
      \right) k {\rm d} k {\rm d } \varphi \notag \\
      &=  \frac{p^2}{2m_e} (1-\beta^2)
     -  \int\limits_0^\infty 
     \frac{e^2  \omega_{\rm t} \hbar (r_0 - r_\infty) }{ (8 \pi  \varepsilon_0)}
     \frac{1}{ \sqrt{ (\varepsilon +  r_0 k ) (\varepsilon +  r_\infty k )} }  \frac{ 1}{ (\varepsilon +  r_\infty k )}
     \frac{ 1}{ \hbar \omega_{{\rm l}, k}
     + \frac{\hbar^2 k^2}{2m_e} }
     \left( 1 + \frac{ \frac{\hbar^2 }{2m_e^2} (1-\beta)^2 ( p k )^2 }{ \left( \hbar \omega_{{\rm l}, k}
     + \frac{\hbar^2 k^2}{2m_e} \right)^2 }
      \right) k {\rm d} k  \notag \\
\end{align}
The first term in integral corresponds to polaron binding energy, and has a form
\begin{align}
     E_{\rm b}   &=  - \frac{e^2  (r_0 - r_\infty) }{8\pi  \varepsilon_0} 
     \int\limits_0^\infty     \frac{1}{ (\varepsilon +  r_0 k ) (\varepsilon +  r_\infty k) }   \frac{k {\rm d} k}{1+ r_{e, {\rm t}}^2 k^2  \sqrt{\frac{\varepsilon +  r_\infty k }{\varepsilon +  r_0 k }} }.
\end{align}
Similarly, the second term has a form
\begin{align}
    E_2 = -  \frac{e^2   (r_0 - r_\infty) }{ (8 \pi  \varepsilon_0) (\hbar\omega_{\rm t})^2 } \frac{\hbar^2 }{2m_e^2} (1-\beta)^2 p^2
     \int\limits_0^\infty 
    \frac{ 1}{ (\varepsilon +  r_0 k )^2}
    \frac{  k^3 {\rm d} k}{ \left( 1   + \frac{\hbar k^2}{2 \omega_{\rm t} m_e} \sqrt{\frac{\varepsilon +  r_\infty k}{\varepsilon +  r_0 k}} \right)^3 }
        \notag \\
\end{align}

Introducing dimensionless quantities
\begin{align}
    r_\infty k /\varepsilon = \Tilde{k}, \quad 
    r_0 / r_\infty = \sigma_0, \quad 
    \varepsilon r_{e, {\rm t}} / r_\infty = \sigma_{e, {\rm t}} ,
\end{align}
we can rewrite
\begin{align}
    E &= - \frac{e^2 }{4\pi \varepsilon_0 r_\infty} \frac{\sigma_0-1}{2}
     \int\limits_0^\infty     \frac{1}{ (1 +  \sigma_0 \Tilde{k} ) (1 +  \Tilde{k}) }   \frac{\Tilde{k} {\rm d} \Tilde{k}}{1+ \sigma_{e, {\rm t}}^2 \Tilde{k}^2  \sqrt{\frac{1 + \Tilde{k} }{1 + \sigma_0 \Tilde{k} }} } \notag \\
     &+\frac{p^2}{2m_e} \left[ (1-\beta^2)
      -  \frac{e^2 }{4\pi  \varepsilon_0 r_\infty} \frac{\sigma_0-1}{2}
      \frac{1 }{ (r_\infty \hbar\omega_{\rm t})^2 } \frac{\hbar^2 }{m_e} (1-\beta)^2 
     \int\limits_0^\infty 
    \frac{ 1}{ (1 +  \sigma_0 \Tilde{k} )^2}
    \frac{  \Tilde{k}^3 {\rm d} \Tilde{k}}{ \left( 1   + \sigma_{e, {\rm t}}^2 \Tilde{k}^2 \sqrt{\frac{1+  \Tilde{k}}{1 +  \sigma_0 \Tilde{k}}} \right)^3 } \right] \notag\\
    &= -E_0 I_{\rm b} + \frac{p^2}{2m_e} \left[ (1-\beta^2) -  \frac{E_0}{\hbar \omega_{\rm t}} I_m (1-\beta)^2 \right],
\end{align}
where
\begin{align}
    E_0 =\frac{e^2 }{4\pi \varepsilon_0 r_\infty} ,
\end{align}
\begin{align}
     I_{\rm b} &= 
     \int\limits_0^\infty     \frac{\sigma_0-1}{2 (1 +  \sigma_0 \Tilde{k} ) (1 +  \Tilde{k}) }   \frac{\Tilde{k} {\rm d} \Tilde{k}}{1+ \sigma_{e, {\rm t}}^2 \Tilde{k}^2  \sqrt{\frac{1 +  \Tilde{k} }{1 +  \sigma_0 \Tilde{k} }} }, \\
     I_m &= 
     \int\limits_0^\infty     \frac{\sigma_{e, {\rm t}}^2 (\sigma_0-1)}{ (1 +  \sigma_0 \Tilde{k} )^2 }   \frac{\Tilde{k}^3 {\rm d} \Tilde{k}}{\left( 1+ \sigma_{e, {\rm t}}^2 \Tilde{k}^2  \sqrt{\frac{1 +  \Tilde{k} }{1 +  \sigma_0 \Tilde{k} }} \right)^3 } .
\end{align}
Similarly, we rewrite
\begin{align}
    \beta 
     &= \frac{e^2 }{4\pi \varepsilon_0 r_\infty} \frac{\sigma_0-1}{2} 
     \frac{(1-\beta)}{m_e \omega_{\rm t}^2r_\infty^2}
     \int\limits_0^\infty     \frac{1}{ (1 +  \sigma_0 \Tilde{k} )^2 }   \frac{\Tilde{k}^3 {\rm d} \Tilde{k}}{\left( 1+ \sigma_{e, {\rm t}}^2 \Tilde{k}^2  \sqrt{\frac{1 +  \Tilde{k} }{1 +  \sigma_0 \Tilde{k} }} \right)^3 }  
     =\frac{E_0}{\hbar \omega_{\rm t}} I_m (1-\beta).
\end{align}
Denoting 
\begin{align}
    \chi_e = \frac{E_0}{\hbar \omega_{\rm t}} 
    I_m (\sigma_0, \sigma_{e, {\rm t}}),
\end{align}
we get 
\begin{align}
    \beta = \frac{\chi_e}{1+\chi_e},
\end{align}
and
\begin{align}
    E  &= -E_0 I_{\rm b} (\sigma_0, \sigma_{e, {\rm t}}) + \frac{p^2}{2m_e} \frac{1}{1+\chi_e }.
\end{align}
We can further proceed with approximate expressions for the integrals $I_{\rm b}$, $I_m$.  One can note, that at the limit $\tilde{k} \gg 1$ one  has $\sqrt{(1+\Tilde{k})/(1+\sigma_0\Tilde{k})} \approx 1/\sqrt{\sigma_0}$.
We therefore perform the corresponding substitution, 
and get an expression which can be readily integrated, resulting in 
\begin{figure}[b]
    \centering
    \includegraphics[width=0.5\linewidth]{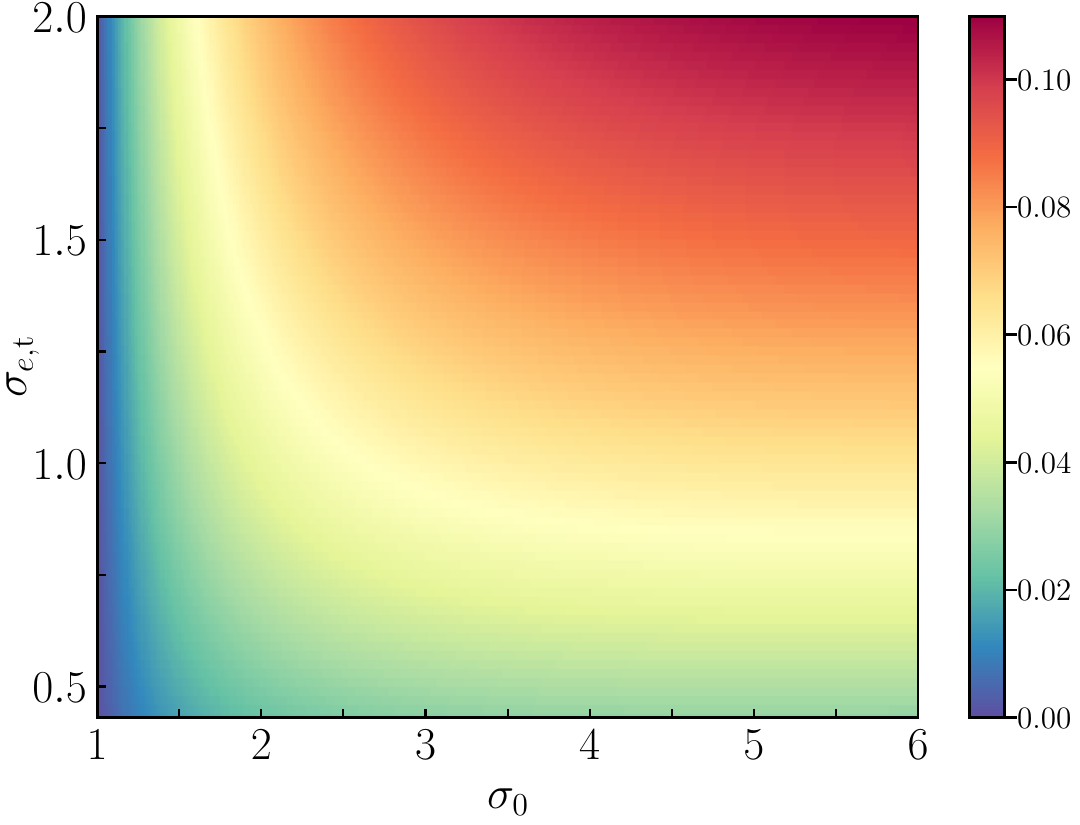}
    \caption{The relative error of polaron binding energy integral and its analytic estimate $( I_{\rm b} -   I_{\rm b}^{\rm approx}) /  I_{\rm b}$. }
    \label{fig:error}
\end{figure}

\begin{align}
    I_{\rm b}^{\rm approx} =  \frac{\sigma_0-1}{2(\sigma_0^2+\Tilde{\sigma}_{e, {\rm t}}^2)}
    \left( \frac{\pi  (\sigma_0+1) \Tilde{\sigma}_{e, {\rm t}} }{2 \left(\Tilde{\sigma}_{e, {\rm t}}^2+1\right)}
    +\frac{\left(\Tilde{\sigma}_{e, {\rm t}}^2-\sigma_0\right) \log \Tilde{\sigma_{e, {\rm t}} } }{\Tilde{\sigma}_{e, {\rm t}}^2+1}
    -\frac{\sigma_0 \log \sigma_0}{\sigma_0-1} \right),
\end{align}
where $\Tilde{\sigma}_{e, {\rm t}} = \sigma_{e, {\rm t}} / \sigma_0^{1/4}$.
Fig.~\ref{fig:error} shows the relative error defined
as $( I_{\rm b} -   I_{\rm b}^{\rm approx}) /  I_{\rm b}$, demonstrating that the approximation works well at the limit $\sigma_0 -1 \ll 1$.
Applying the same approximation, we get
\begin{align}
   I_m^{\rm approx} =
   \frac{\sigma_{e, {\rm t}}^2(\sigma_0-1)}{8\tilde{\sigma}_{e,{\rm t}}^2(\sigma_0^2+\tilde{\sigma}_{e,{\rm t}}^2)^4} \times \notag \\ 
   \left[ 
   2(\sigma_0^6+ \tilde{\sigma}_{e,{\rm t}}^6)
- 3\pi \sigma_0\tilde{\sigma}_{e,{\rm t}} \left( \sigma_0^4
+\tilde{\sigma}_{e,{\rm t}}^4 \right)
- 18\sigma_0^2\tilde{\sigma}_{e,{\rm t}}^2 (\sigma_0^2 + \tilde{\sigma}_{e,{\rm t}}^2 -\pi)
+ 24\sigma_0^2 \tilde{\sigma}_{e,{\rm t}}^2 (\sigma_0^2-\tilde{\sigma}_{e,{\rm t}}^2)
\log\!\biggl(\sigma_0/\tilde{\sigma}_{e,{\rm t}}\biggr) 
\right] .
\end{align}
\subsection{Landau-Pekar model}

The polaron binding energy within the Landau-Pekar model can be found via the variational minimization of the following functional \cite{devreese2016fr}:
\begin{align}
    E_e^{\rm LP} 
    = \left\langle \psi \left| \frac{\hat{p}_e^2}{2m_e} \right| \psi \right\rangle
    - \sum_k \frac{ |V_k|^2 |g_k|^2 }{\hbar \omega_{{\rm l},k} }, 
\end{align}
\begin{table}[t]
\caption{Binding energy of electron- and hole- polarons calculated calculated via different methods.
Here $E_e^{\rm b}$ is the energy of Fr{\"oh}lich electron-polaron, $E_e^{\rm LP}$ is the Landau-Pekar binding energy, calculated via Eq.~\eqref{eq:EeLP},
and $E_e^{\rm LP*}$ is the Landau-Pekar binding energy calculated within the trivial limit of LO phonon dispersion in Ref.~\cite{sio2023polarons}.
Similar results are shown for hole-polarons.}
\label{table:S1}
\begin{ruledtabular}
\begin{tabular}{lcccccc}
Material & $E_e^{\rm b}$ (meV) & $E_e^{\rm LP}$ (meV) & $E_e^{\rm LP*}$ (meV) & $E_h^{\rm b}$ (meV) & $E_h^{\rm LP}$ (meV) & $E_h^{\rm LP*}$ (meV) \\[3pt]\hline
h-BN     & 129.3   & 12.00 &  17.5 & 116.3   & 7.405 &  10.8\\
GaN      & 55.16   & 0.460  & 0.7 & 132.0   & 37.47 & 57.4 \\
AlN      & 105.4   & 9.607 & 17.2 & 186.5   & 66.83  &  119.9\\
HfSe$_2$ & 89.56   & 64.05 & 402.4 & 98.51   & 73.72  &  458.0\\
HfS$_2$  & 123.2   & 82.04  & 469.0 & 155.1   & 115.8 &  646.0\\
ZrS$_2$  & 142.9   & 80.02 & 349.0  & 133.7   & 71.34  & 313.1 \\
\end{tabular}
\end{ruledtabular}
\end{table}
where $g_k = \langle \psi \left| e^{i \mathbf{k} \cdot \mathbf{r} } \right| \psi \rangle$. 
The trial function is chosen in the  2D hydrogen-like form: 
\begin{align}
    \psi = \sqrt{\frac{2}{\pi}} \frac{1}{a} e^{-r/a},
\end{align}
where $a$ is a variational parameter.
The kinetic energy is then readily evaluated as
\begin{align}
     \left\langle \psi \left| \frac{\hat{p}_e^2}{2m_e} \right| \psi \right\rangle = \frac{\hbar^2}{2m_e a^2},
\end{align}
and 
\begin{align}
    g_k = \frac{1}{\left( 1+ (ak)^2/4 \right)^{3/2}} .
\end{align}
The explicit expression for the effective potential is
\begin{align}
     \frac{ |V_k|^2 }{\hbar \omega_{{\rm l},k} } =
     \frac{ 1  }{\hbar \omega_{{\rm l},k} } 
    \frac{e^2  \omega_{\rm t}^2 \hbar (r_0 - r_\infty) }{ 4 \varepsilon_0 A \omega_{{\rm l},k} }
    \frac{ 1}{ (\varepsilon +  r_\infty k )^2}
    =  \frac{e^2 (r_0 - r_\infty) }{ 4 \varepsilon_0 A  }
    \frac{ 1}{ (\varepsilon +  r_\infty k )(\varepsilon +  r_0 k )}.
\end{align}

Then one has
\begin{align}
    \label{eq:EeLP}
    E_e^{\rm LP} 
    &= \frac{\hbar^2}{2m_e a^2}
    - \frac{A}{(2\pi)^2} \int\limits_0^\infty \frac{e^2   (r_0 - r_\infty) }{ 4 \varepsilon_0 A  }
    \frac{ 1}{ (\varepsilon +  r_\infty k )}
    \frac{ 1}{ (\varepsilon +  r_0 k )}
    \frac{2\pi k {\rm d}k  }{\left( 1+ (ak)^2/4 \right)^3} \notag \\
    &= \frac{\hbar^2}{2m_e a^2}
    - \frac{e^2}{4\pi \varepsilon_0 r_\infty} \frac{\sigma_0-1}{2}
    \int\limits_0^\infty  
    \frac{ 1}{ (1+  \tilde{k})}
    \frac{ 1}{ (1 +  \sigma_0 \tilde{k} )}
    \frac{ \tilde{k} {\rm d} \tilde{k}  }{\left( 1+  \left( \tilde{k} \frac{ \varepsilon a}{2 r_\infty} \right) ^2  \right)^3}.
\end{align}
Variational minimization of the last expression gives the polaron binding energy in the Landau-Pekar treatment. 
The results are shown in Table~\ref{table:S1}.
One can find essentially smaller binding energy obtained within Landau-Pekar approach, in line with the 3D case \cite{kittel1987quantum}. 
This difference is especially dramatic for weakly polar materials, where $r_0 - r_\infty < r_\infty$, such as h-BN, GaN, AlN.
We additionally address the results of  Ref.~\cite{sio2023polarons}, where the effective potential for Landau-Pekar model is 
\begin{align}
    \frac{V_k^2}{ \hbar \omega}=
    \frac{e^2 (\epsilon_0 -\epsilon_\infty) d q_c^2 }{16 \pi \varepsilon_0} \frac{2\pi}{A} \frac{1}{(k+q_c)^2},
\end{align}
with $q_c = 1/[ d \epsilon_\infty/2 - d / (4\epsilon_\infty) ] $.
Substituting $d \epsilon_\infty/2 = r_\infty$, $d \epsilon_0/2 = r_0$, we get 
$q_c = 1/[ r_\infty - r_\infty / (2\epsilon_\infty^2) ] $, and
\begin{align}
    \frac{V_k^2}{ \hbar \omega}=
    \frac{e^2 (r_0 -r_\infty)  }{4 \varepsilon_0 A}  \frac{1}{[1 + k r_\infty (1- 1 /(2\epsilon_\infty^2) ) ]^2}.
\end{align}
Now, taking into account the dielectric constant of surrounding media $\varepsilon=1$, we get
\begin{align}
    \label{eq:EeLP2}
    E_e^{\rm LP*} 
    &= \frac{\hbar^2}{2m_e a^2}
    - \frac{A}{(2\pi)^2} \int\limits_0^\infty \frac{e^2   (r_0 - r_\infty) }{ 4 \varepsilon_0 A  }
    \frac{1}{[1 + k r_\infty (1- 1 /(2\epsilon_\infty^2) ) ]^2}
    \frac{2\pi k {\rm d}k  }{\left( 1+ (ak)^2/4 \right)^3} \notag \\
    &= \frac{\hbar^2}{2m_e a^2}
    - \frac{e^2}{4\pi \varepsilon_0 r_\infty} \frac{\sigma_0-1}{2}
    \int\limits_0^\infty  
    \frac{ 1}{ \left[ 1+  \tilde{k} (1- 1 /(2\epsilon_\infty^2) )  \right]^2}
    \frac{ \tilde{k} {\rm d} \tilde{k}  }{\left( 1+  \left( \tilde{k} \frac{  a}{2 r_\infty} \right) ^2  \right)^3}.
\end{align}
The variational minimization of Eq.~\eqref{eq:EeLP2} allows to exactly reproduce polaron binding energies reported in Ref.~\cite{sio2023polarons}.
One can thus see that neglecting the dispersion of 2D LO phonons has a dramatic impact on polaron binding energy, leading to overestimation for strongly polar materials with $r_0 \gg r_\infty$.


\section{Exciton states}

We next consider the exciton-polaron problem.
The Hamiltonian of an electron-hole pair interacting with LO phonons can be written as
\begin{align}
    \hat{H}_{\rm X} &= \frac{\hat{\mathbf{p}}_e^2}{2m_e} + \frac{\hat{\mathbf{p}}_h^2}{2m_h}
    + V_{\rm KR} (|\mathbf{r}_e - \mathbf{r}_h|,r_\infty)
    + \sum_k \hbar \omega_{{\rm l}, k} 
    \hat{a}_\mathbf{k}^\dag \hat{a}_\mathbf{k} \notag \\
    &+ \sum_k  \left( V_k e^{i\mathbf{k}\cdot\mathbf{r}_e} \hat{a}_{\mathbf{k}}  +  V_k^* e^{-i\mathbf{k}\cdot\mathbf{r}_e} \hat{a}_{\mathbf{k}}^\dag \right)
    -\sum_k  \left( V_k e^{i\mathbf{k}\cdot\mathbf{r}_h} \hat{a}_{\mathbf{k}}  +  V_k^* e^{-i\mathbf{k}\cdot\mathbf{r}_h} \hat{a}_{\mathbf{k}}^\dag \right),
\end{align}
where $V_{\rm KR} $ is the bare 2D Coulomb potential. 
Proceeding to the center-of-mass and relative coordinates,
one can rewrite
\begin{align}
    \hat{H}_{\rm X} = \frac{\hat{\mathbf{P}}^2}{2M} + \frac{\hat{\mathbf{p}}^2}{2\mu}
    + V_{\rm KR} (r,r_\infty)
    + \sum_k \hbar \omega_{{\rm l}, k} 
    \hat{a}_\mathbf{k}^\dag \hat{a}_\mathbf{k} 
    + \sum_k  \left( V_k \rho_{\mathbf{k}} e^{i\mathbf{k}\cdot\mathbf{R}} \hat{a}_{\mathbf{k}}
    +  V_k^* \rho_{\mathbf{k}}^* e^{-i\mathbf{k}\cdot\mathbf{R}} \hat{a}_{\mathbf{k}}^\dag \right),
\end{align}
where
\begin{align}
    \rho_{\mathbf{k}} = e^{i m_h \mathbf{k}\cdot\mathbf{r} /M} - e^{-i m_e \mathbf{k}\cdot\mathbf{r} /M}. 
\end{align}
We perform an unitary transform with the following operator
\begin{align}
    \hat{U}_1 = e^{ i (\mathbf{Q} - \sum_k \mathbf{k} 
    \hat{a}_\mathbf{k}^\dag \hat{a}_\mathbf{k}) \cdot \mathbf{R}}, 
\end{align}
where $\hbar \mathbf{Q}$ corresponds to overall center-of-mass momentum.
Similarly to the case of single polaron, we get the following relations
\begin{align}
    \hat{U}_1^{-1} \hat{\mathbf{P}} \hat{U}_1 &=  \hat{\mathbf{P}}  + \hbar \mathbf{Q} 
    - \hbar \sum_k \hat{a}_\mathbf{k}^\dag \hat{a}_\mathbf{k} \mathbf{k},  \\
    \hat{U}_1^{-1} \hat{\mathbf{p}} \hat{U}_1 &=  \hat{\mathbf{p}}, \\ 
    \hat{U}_1^{-1} \hat{a}_{\mathbf{k}} \hat{U}_1 &=  \hat{a}_{\mathbf{k}}  e^{-i \mathbf{k} \cdot \mathbf{R}},
\end{align}
yielding in the transformed Hamiltonian
\begin{align}
    \Tilde{\hat{H}}_{\rm X} =  \frac{(\hat{\mathbf{P}}  + \hbar \mathbf{Q} 
    - \hbar \sum_k \hat{a}_\mathbf{k}^\dag \hat{a}_\mathbf{k} \mathbf{k})^2}{2M} + \frac{\hat{\mathbf{p}}^2}{2\mu}
    + V_{\rm KR} (r,r_\infty)
    + \sum_k \hbar \omega_{{\rm l}, k} 
    \hat{a}_\mathbf{k}^\dag \hat{a}_\mathbf{k} 
    + \sum_k  \left( V_k \rho_{\mathbf{k}} \hat{a}_{\mathbf{k}}
    +  V_k^* \rho_{\mathbf{k}}^*  \hat{a}_{\mathbf{k}}^\dag \right).
\end{align}
We get rid of $\mathbf{R}$-coordinate in Hamiltonian, so that we may set $\hat{\mathbf{P}}  + \hbar \mathbf{Q} \rightarrow \hbar \mathbf{Q}$. 
Expanding the square term of Hamiltonian, we get
\begin{align}
    \Tilde{\hat{H}}_{\rm X} &=  \frac{\hbar^2 \mathbf{Q}^2}{2M}
    - \frac{\hbar^2}{M} \mathbf{Q} \cdot \sum_k \hat{a}_\mathbf{k}^\dag \hat{a}_\mathbf{k} \mathbf{k}
    + \frac{\hbar^2}{2M} \left( \sum_k \hat{a}_\mathbf{k}^\dag \hat{a}_\mathbf{k} \mathbf{k} \right)^2 
    + \frac{\hat{\mathbf{p}}^2}{2\mu}
    + V_{\rm KR} (r,r_\infty)
    + \sum_k \hbar \omega_{{\rm l}, k} 
    \hat{a}_\mathbf{k}^\dag \hat{a}_\mathbf{k} 
    + \sum_k  \left( V_k \rho_{\mathbf{k}} \hat{a}_{\mathbf{k}}
    +  V_k^* \rho_{\mathbf{k}}^*  \hat{a}_{\mathbf{k}}^\dag \right) \notag \\
    &=  \frac{\hbar^2 \mathbf{Q}^2}{2M}
    + \frac{\hat{\mathbf{p}}^2}{2\mu}
    + V_{\rm KR} (r,r_\infty)
    + \sum_k \left( \hbar \omega_{{\rm l}, k} - \frac{\hbar^2}{M} \mathbf{Q} \cdot \mathbf{k} + \frac{\hbar^2 k^2}{2M}
    \right)
    \hat{a}_\mathbf{k}^\dag \hat{a}_\mathbf{k} 
    + \frac{\hbar^2}{2M}  \sum_k \sum_q \mathbf{k}  
    \cdot    \mathbf{q}
    \hat{a}_\mathbf{k}^\dag \hat{a}_\mathbf{q}^\dag \hat{a}_\mathbf{q} \hat{a}_\mathbf{k} \notag \\
    &+ \sum_k  \left( V_k \rho_{\mathbf{k}} \hat{a}_{\mathbf{k}}
    +  V_k^* \rho_{\mathbf{k}}^*  \hat{a}_{\mathbf{k}}^\dag \right).
\end{align}
We proceed with another transformation of Hamiltonian with unitary operator
\begin{align}
    \hat{U}_2 = e^{ \sum_k ( \hat{a}_\mathbf{k} F^*_{\mathbf{k}} (\mathbf{r}) - \hat{a}_\mathbf{k}^\dag F_{\mathbf{k}} (\mathbf{r})  ) }
\end{align}
where $F_{\mathbf{k}} (\mathbf{r})$ being the mean field value of the displacement.
We then have
\begin{align}
    \hat{U}_2^{-1} \hat{a}_{\mathbf{k}} \hat{U}_2 = \hat{a}_{\mathbf{k}} + \left[\sum_q (- \hat{a}_\mathbf{q} F^*_{\mathbf{q}} (\mathbf{r}) + \hat{a}_\mathbf{q}^\dag F_{\mathbf{q}} (\mathbf{r})  ), \hat{a}_{\mathbf{k}} \right] + \cdots 
    = \hat{a}_{\mathbf{k}} - \sum_q \delta_{\mathbf{q},\mathbf{k}} F_{\mathbf{q}} (\mathbf{r})
    =\hat{a}_{\mathbf{k}} - F_{\mathbf{k}} (\mathbf{r}), 
\end{align}
\begin{align}
    \hat{U}_2^{-1} \hat{a}_{\mathbf{k}}^\dag \hat{U}_2 = \hat{a}_{\mathbf{k}}^\dag - F_{\mathbf{k}}^* (\mathbf{r}),
\end{align}
and
\begin{align}
    &\hat{U}_2^{-1} \hat{\mathbf{p}} \hat{U}_2 = \hat{\mathbf{p}} 
    +\left[\sum_k (- \hat{a}_\mathbf{k} F^*_{\mathbf{k}} (\mathbf{r}) + \hat{a}_\mathbf{k}^\dag F_{\mathbf{k}} (\mathbf{r})  ), \hat{\mathbf{p}} \right]
    +\frac{1}{2} \left[\sum_{q} (- \hat{a}_{\mathbf{q}} F^*_{\mathbf{q}} (\mathbf{r}) + \hat{a}_{\mathbf{q}}^\dag F_{\mathbf{q}} (\mathbf{r})  ), 
    \left[\sum_k (- \hat{a}_\mathbf{k} F^*_{\mathbf{k}} (\mathbf{r}) + \hat{a}_\mathbf{k}^\dag F_{\mathbf{k}} (\mathbf{r})  ), \hat{\mathbf{p}} \right] \right] + \cdots \notag \\
    &= \hat{\mathbf{p}} 
    -2\mu \frac{\hbar}{2i\mu} \sum_k \left( \hat{a}_\mathbf{k}^\dag \nabla F_{\mathbf{k}} (\mathbf{r}) -  \hat{a}_\mathbf{k} \nabla F^*_{\mathbf{k}} (\mathbf{r})     \right)
    +\mu \frac{\hbar}{2i\mu} \sum_{k} \left( F^*_{\mathbf{k}} (\mathbf{r}) \nabla F_{\mathbf{k}} (\mathbf{r}) 
    -  F_{\mathbf{k}} (\mathbf{r}) \nabla F^*_{\mathbf{k}} (\mathbf{r}) \right)    
    =\hat{\mathbf{p}} -2\mu \mathbf{J} (\mathbf{r}, \hat{a}_{\mathbf{k}}) + \mu \mathbf{j} (\mathbf{r}) .
\end{align}
The Hamiltonian terms are transformed as follows:
\begin{align}
    &\hat{U}_2^{-1} \left( \frac{\hat{\mathbf{p}}^2}{2\mu}
   \right) \hat{U}_2 
    = \frac{1}{2\mu} \hat{U}_2^{-1} \hat{\mathbf{p}} \hat{U}_2 \cdot \hat{U}_2^{-1} \hat{\mathbf{p}} \hat{U}_2
    = \frac{1}{2\mu} \left ( \hat{\mathbf{p}} -2\mu \mathbf{J} (\mathbf{r}, \hat{a}_{\mathbf{k}}) + \mu \mathbf{j} (\mathbf{r}) \right)^2 \notag \\
    &= \frac{1}{2\mu} \left ( \hat{\mathbf{p}}^2 
    +\mu^2 \mathbf{j}^2 + \mu ( \hat{\mathbf{p}} \cdot \mathbf{j} + \mathbf{j} \cdot \hat{\mathbf{p}})
    +4\mu^2 \mathbf{J}^2 
    -4\mu^2 \mathbf{J} \cdot \mathbf{j}  
    -2\mu ( \hat{\mathbf{p}} \cdot \mathbf{J} + \mathbf{J} \cdot \hat{\mathbf{p}}) \right) \notag \\
    &= \frac{\left ( \hat{\mathbf{p}} 
    +\mu \mathbf{j} \right)^2}{2\mu} 
    +\frac{\hbar^2}{2\mu}  \sum_k \nabla F^*_{\mathbf{k}} \cdot \nabla F_{\mathbf{k}}
    +\frac{\hbar^2}{2\mu}  \sum_k \sum_q 
    \left(2\hat{a}_\mathbf{k}^\dag \hat{a}_\mathbf{q}
    \nabla F_{\mathbf{k}} \cdot \nabla F^*_{\mathbf{q}}
    -\hat{a}_\mathbf{k}^\dag \hat{a}_\mathbf{q}^\dag \nabla F_{\mathbf{k}}  \cdot \nabla F_{\mathbf{q}} 
    -  \hat{a}_\mathbf{k} \hat{a}_\mathbf{q} \nabla F^*_{\mathbf{k}} \cdot \nabla F^*_{\mathbf{q}}  
    \right)  \notag \\
    &+\frac{\hbar}{i\mu} \left( \hat{a}_\mathbf{k} \nabla F^*_{\mathbf{k}} - \hat{a}_\mathbf{k}^\dag \nabla F_{\mathbf{k}} \right) \cdot \left( \hat{\mathbf{p}} + \mu \mathbf{j} \right)
    + \frac{\hbar^2}{2\mu} \left( \hat{a}_\mathbf{k}^\dag \Delta F_{\mathbf{k}} -  \hat{a}_\mathbf{k} \Delta F^*_{\mathbf{k}}     \right), 
\end{align}
\begin{align}
    \hat{U}_2^{-1} \left( \frac{\hbar^2 \mathbf{Q}^2}{2M}
    + V_{\rm KR} (r,r_\infty) \right) \hat{U}_2 
    &= \frac{\hbar^2 \mathbf{Q}^2}{2M}
    + V_{\rm KR} (r,r_\infty), \\
    \hat{U}_2^{-1} \left( \sum_k \left( \hbar \omega_{{\rm l}, k} - \frac{\hbar^2}{M} \mathbf{Q} \cdot \mathbf{k} + \frac{\hbar^2 k^2}{2M}
    \right)
    \hat{a}_\mathbf{k}^\dag \hat{a}_\mathbf{k}  \right) \hat{U}_2 
    &=\sum_k \left( \hbar \omega_{{\rm l}, k} - \frac{\hbar^2}{M} \mathbf{Q} \cdot \mathbf{k} + \frac{\hbar^2 k^2}{2M}
    \right)
    \left( \hat{a}_\mathbf{k}^\dag \hat{a}_\mathbf{k} + F_\mathbf{k}^* F_\mathbf{k} -\hat{a}_\mathbf{k}^\dag F_\mathbf{k} -\hat{a}_\mathbf{k} F_\mathbf{k}^* \right), \\
    \hat{U}_2^{-1} \left( \sum_k  \left( V_k \rho_{\mathbf{k}} \hat{a}_{\mathbf{k}}
    +  V_k^* \rho_{\mathbf{k}}^*  \hat{a}_{\mathbf{k}}^\dag \right)
    \right) \hat{U}_2 
    &=\sum_k  \left( V_k \rho_{\mathbf{k}} \hat{a}_{\mathbf{k}}
    +  V_k^* \rho_{\mathbf{k}}^*  \hat{a}_{\mathbf{k}}^\dag 
    -V_k \rho_{\mathbf{k}} F_{\mathbf{k}}
    - V_k^* \rho_{\mathbf{k}}^*  F_{\mathbf{k}}^* \right),
\end{align}
\begin{align}
    &\hat{U}_2^{-1} \left( \frac{\hbar^2}{2M}  \sum_k \sum_q \mathbf{k}  
    \cdot    \mathbf{q}
    \hat{a}_\mathbf{k}^\dag \hat{a}_\mathbf{q}^\dag \hat{a}_\mathbf{q} \hat{a}_\mathbf{k} \right) \hat{U}_2 
    =\frac{\hbar^2}{2M}  \sum_k \sum_q \mathbf{k}  
    \cdot    \mathbf{q} 
    \left( \hat{a}_\mathbf{k}^\dag - F_\mathbf{k}^*\right)
    \left( \hat{a}_\mathbf{q}^\dag - F_\mathbf{q}^*\right)
    \left( \hat{a}_\mathbf{q} - F_\mathbf{q}\right)
    \left( \hat{a}_\mathbf{k} - F_\mathbf{k}\right) \notag \\
    &=\frac{\hbar^2}{2M}  \sum_k \sum_q \mathbf{k}  
    \cdot    \mathbf{q} 
    \left( \hat{a}_\mathbf{k}^\dag \hat{a}_\mathbf{q}^\dag \hat{a}_\mathbf{q} \hat{a}_\mathbf{k}
          -\hat{a}_\mathbf{k}^\dag \hat{a}_\mathbf{q}^\dag \hat{a}_\mathbf{q} F_\mathbf{k}
          -\hat{a}_\mathbf{k}^\dag \hat{a}_\mathbf{q}^\dag \hat{a}_\mathbf{k} \hat{F}_\mathbf{q}
          +\hat{a}_\mathbf{k}^\dag \hat{a}_\mathbf{q}^\dag F_\mathbf{q} F_\mathbf{k} 
    \right. \notag \\
    &\hspace{80 pt}   -\hat{a}_\mathbf{k}^\dag F_\mathbf{q}^* \hat{a}_\mathbf{q} \hat{a}_\mathbf{k}
                      +\hat{a}_\mathbf{k}^\dag F_\mathbf{q}^* \hat{a}_\mathbf{q} F_\mathbf{k}
                      +\hat{a}_\mathbf{k}^\dag F_\mathbf{q}^* \hat{a}_\mathbf{k} \hat{F}_\mathbf{q}
                      -\hat{a}_\mathbf{k}^\dag F_\mathbf{q}^* F_\mathbf{q} F_\mathbf{k} \notag \\
    &\hspace{80 pt}   -F_\mathbf{k}^* \hat{a}_\mathbf{q}^\dag \hat{a}_\mathbf{q} \hat{a}_\mathbf{k}
                      +F_\mathbf{k}^* \hat{a}_\mathbf{q}^\dag \hat{a}_\mathbf{q} F_\mathbf{k}
                      +F_\mathbf{k}^* \hat{a}_\mathbf{q}^\dag \hat{a}_\mathbf{k} \hat{F}_\mathbf{q}
                      -F_\mathbf{k}^* \hat{a}_\mathbf{q}^\dag F_\mathbf{q} F_\mathbf{k} \notag \\
&\hspace{80 pt}\left. +F_\mathbf{k}^* F_\mathbf{q}^* \hat{a}_\mathbf{q} \hat{a}_\mathbf{k}
                      -F_\mathbf{k}^* F_\mathbf{q}^* \hat{a}_\mathbf{q} F_\mathbf{k}
                      -F_\mathbf{k}^* F_\mathbf{q}^* \hat{a}_\mathbf{k} \hat{F}_\mathbf{q}
                      +F_\mathbf{k}^* F_\mathbf{q}^* F_\mathbf{q} F_\mathbf{k} \right) .
\end{align}

The transformed Hamiltonian then is
\begin{align}
    \hat{H}_{\rm X} = \hat{H}_{\rm X0} + \hat{H}_{\rm X1} + \hat{H}_{\rm X2} + \hat{H}_{\rm X3} +\hat{H}_{\rm X4},
\end{align}
where
\begin{align}
    \hat{H}_{\rm X0} &= \frac{\hbar^2 \mathbf{Q}^2}{2M} + V_{\rm KR} (r,r_\infty)  
    + \frac{\left ( \hat{\mathbf{p}} +\mu \mathbf{j} \right)^2}{2\mu} 
    +\frac{\hbar^2}{2\mu}  \sum_k \nabla F^*_{\mathbf{k}} \cdot \nabla F_{\mathbf{k}}
    +\sum_k \left( \hbar \omega_{{\rm l}, k} - \frac{\hbar^2}{M} \mathbf{Q} \cdot \mathbf{k} + \frac{\hbar^2 k^2}{2M}
    \right) |F_\mathbf{k}|^2 \notag \\
                     &-\sum_k \left( V_k \rho_{\mathbf{k}} F_{\mathbf{k}} 
    + V_k^* \rho_{\mathbf{k}}^*  F_{\mathbf{k}}^* \right)
    + \frac{\hbar^2}{2M}  \sum_k \sum_q \mathbf{k}  
    \cdot    \mathbf{q} |F_\mathbf{k}|^2 |F_\mathbf{q}|^2  \\
    \hat{H}_{\rm X1} &= \sum_k \left( V_k \rho_{\mathbf{k}} 
    -\left( \hbar \omega_{{\rm l}, k} - \frac{\hbar^2}{M} \mathbf{Q} \cdot \mathbf{k} + \frac{\hbar^2 k^2}{2M} 
    + \frac{\hbar^2}{M} \sum_q \mathbf{k}  \cdot \mathbf{q} |F_{\mathbf{q}}|^2 \right) F_\mathbf{k}^* 
    - \frac{\hbar^2}{2\mu} \Delta F^*_{\mathbf{k}}
    +\frac{\hbar}{i\mu} \nabla F_{\mathbf{k}}^* \cdot \left( \hat{\mathbf{p}} + \mu \mathbf{j} \right)
    \right) \hat{a}_\mathbf{k} \notag \\
                     &+ \sum_k \left( V_k^* \rho_{\mathbf{k}}^* 
                     -\left( \hbar \omega_{{\rm l}, k} - \frac{\hbar^2}{M} \mathbf{Q} \cdot \mathbf{k} + \frac{\hbar^2 k^2}{2M} 
                     + \frac{\hbar^2}{M} \sum_q \mathbf{k}  \cdot \mathbf{q} |F_{\mathbf{q}}|^2 \right) F_\mathbf{k}
                     + \frac{\hbar^2}{2\mu} \Delta F_{\mathbf{k}}
                     - \frac{\hbar}{i\mu} \nabla F_{\mathbf{k}} \cdot \left( \hat{\mathbf{p}} + \mu \mathbf{j} \right)
                     \right) \hat{a}_\mathbf{k}^\dag \notag \\
    \hat{H}_{\rm X2} &= \sum_k \left( \hbar \omega_{{\rm l}, k} - \frac{\hbar^2}{M} \mathbf{Q} \cdot \mathbf{k} + \frac{\hbar^2 k^2}{2M} 
    + \frac{\hbar^2}{M} \sum_q \mathbf{k}  \cdot \mathbf{q} |F_{\mathbf{q}}|^2 \right)
    \hat{a}_\mathbf{k}^\dag \hat{a}_\mathbf{k} \notag \\
                    &+ \sum_k \sum_q \left( \frac{\hbar^2}{M} \mathbf{k}  \cdot  \mathbf{q} F_\mathbf{q}^* F_\mathbf{k} 
                    +\frac{\hbar^2}{\mu} \nabla F_{\mathbf{k}} \cdot \nabla F^*_{\mathbf{q}}  \right) 
                    \hat{a}_\mathbf{k}^\dag \hat{a}_\mathbf{q} \notag \\
                    &+ \sum_k \sum_q \left( \frac{\hbar^2}{2M} \mathbf{k}  \cdot  \mathbf{q} F_\mathbf{q}^* F_\mathbf{k}^* 
                    -\frac{\hbar^2}{2\mu} \nabla F_{\mathbf{k}}^* \cdot \nabla F^*_{\mathbf{q}}  \right) 
                    \hat{a}_\mathbf{k} \hat{a}_\mathbf{q} \notag \\
                    &+ \sum_k \sum_q \left( \frac{\hbar^2}{2M} \mathbf{k}  \cdot  \mathbf{q} F_\mathbf{q} F_\mathbf{k} 
                    -\frac{\hbar^2}{2\mu} \nabla F_{\mathbf{k}} \cdot \nabla F_{\mathbf{q}}  \right) 
                    \hat{a}_\mathbf{k}^\dag \hat{a}_\mathbf{q}^\dag \\
    \hat{H}_{\rm X3} &=  -\frac{\hbar^2}{M}  \sum_k \sum_q \mathbf{k}  \cdot  \mathbf{q} 
    \left( \hat{a}_\mathbf{q}^\dag  \hat{a}_\mathbf{k}^\dag \hat{a}_\mathbf{k} \hat{F}_\mathbf{q}
         + \hat{a}_\mathbf{k}^\dag \hat{a}_\mathbf{k} \hat{a}_\mathbf{q} \hat{F}_\mathbf{q}^* \right)  \\
    \hat{H}_{\rm X4} &= \frac{\hbar^2}{2M}  \sum_k \sum_q \mathbf{k}  \cdot  \mathbf{q}
    \hat{a}_\mathbf{k}^\dag \hat{a}_\mathbf{q}^\dag \hat{a}_\mathbf{q} \hat{a}_\mathbf{k}
\end{align}
In analogy to single polaron problem, we have $\hat{a}_{\mathbf{k}} \psi_{\rm X} =0$, where $\psi_{\rm X} =0$ is the exciton wave function. 
Then for the variational energy functional one has $E_{\rm X} = \langle \psi_{\rm X} |\hat{H}_{\rm X}|\psi_{\rm X} \rangle
= \langle \psi_{\rm X} |\hat{H}_{\rm X0}|\psi_{\rm X} \rangle$.
The ground state exciton wave function ($s$-orbital) assumed to be real, which results in the following relation
\begin{align}
   &\langle \psi_{\rm X} | \hat{p}\cdot \mathbf{j} + \mathbf{j} \cdot \hat{p} |\psi_{\rm X} \rangle
   = -i \hbar \int \psi_{\rm X} \left( 2\mathbf{j} \cdot \nabla \psi_{\rm X} 
   + (\nabla \mathbf{j}) \psi_{\rm X}
   \right) {\rm d}^2 r 
   = -i \hbar \int \left( 2 \psi_{\rm X} \mathbf{j} \cdot \nabla \psi_{\rm X} 
   + (\nabla \mathbf{j}) \psi^2_{\rm X}
   \right) {\rm d}^2 r \notag \\
   &= -i \hbar \int \left( 2 \psi_{\rm X} \mathbf{j} \cdot \nabla \psi_{\rm X} 
   + \nabla ( \mathbf{j} \psi^2_{\rm X} ) 
   - \mathbf{j} \cdot \nabla (\psi^2_{\rm X})
   \right) {\rm d}^2 r 
   = -i \hbar \int \left( 2 \psi_{\rm X} \mathbf{j} \cdot \nabla \psi_{\rm X} 
   - \mathbf{j} \cdot 2 \psi_{\rm X} \nabla \psi_{\rm X}
   \right) {\rm d}^2 r =0
\end{align}
Setting the center-of-mass momentum $\mathbf{Q}=0$, we then get
\begin{align}
    E_{\rm X} &= \langle \psi_{\rm X} |
    V_{\rm KR} (r,r_\infty)  +\frac{ \hat{\mathbf{p}}^2}{2\mu}
    +\frac{\hbar^2}{2\mu}  \sum_k \nabla F^*_{\mathbf{k}} \cdot \nabla F_{\mathbf{k}}
    +\sum_k \left( \hbar \omega_{{\rm l}, k}  + \frac{\hbar^2 k^2}{2M}
    \right) |F_\mathbf{k}|^2 
    -\sum_k \left( V_k \rho_{\mathbf{k}} F_{\mathbf{k}} 
    + V_k^* \rho_{\mathbf{k}}^*  F_{\mathbf{k}}^* \right) |\psi_{\rm X} \rangle , 
\end{align}
where we retain the terms up to $|F|^2$.
We now proceed with an approximate solution, neglecting the mixing of displacement $F_{\mathbf{k}}$ and exciton wave function $\psi_{\rm X}$. 
Then one can minimize the following functional:
\begin{align}
    W_{\rm X} &= \frac{\hbar^2}{2\mu}  \sum_k  \nabla F^*_{\mathbf{k}} \cdot \nabla F_{\mathbf{k}}
    +\sum_k \left( \hbar \omega_{{\rm l}, k}  + \frac{\hbar^2 k^2}{2M}
    \right) |F_\mathbf{k}|^2 
    -\sum_k \left( V_k \rho_{\mathbf{k}} F_{\mathbf{k}} 
    + V_k^* \rho_{\mathbf{k}}^*  F_{\mathbf{k}}^* \right),
\end{align}
\begin{align}
    \frac{\delta W_{\rm X}}{\delta F_{\mathbf{k}}}
    =\frac{\delta W_{\rm X}}{\delta F^*_{\mathbf{k}}} =0.
\end{align}
Evaluating, we get
\begin{align}
    &-\frac{\hbar^2}{2\mu}   \Delta F_{\mathbf{k}}
    + \left( \hbar \omega_{{\rm l}, k}  + \frac{\hbar^2 k^2}{2M}
    \right) F_\mathbf{k} 
    -V_k^* \rho_{\mathbf{k}}^*  =0.
\end{align}
We employ a variational anzats
\begin{align}
    F_{\mathbf{k}} = A_{\mathbf{k}} e^{-i m_h \mathbf{k} \cdot \mathbf{r} /M }
                    -B_{\mathbf{k}} e^{ i m_e \mathbf{k} \cdot \mathbf{r} /M }
\end{align}
yielding in
\begin{align}
    \nabla F_{\mathbf{k}} & =-i \mathbf{k}  \frac{m_h}{M} A_{\mathbf{k}} e^{-i m_h \mathbf{k} \cdot \mathbf{r} /M } 
                    -i \mathbf{k} \frac{m_e}{M} B_{\mathbf{k}} e^{ i m_e \mathbf{k} \cdot \mathbf{r} /M } , \\
    \Delta F_{\mathbf{k}} & = - k^2  \frac{m_h^2}{M^2} A_{\mathbf{k}} e^{-i m_h \mathbf{k} \cdot \mathbf{r} /M } 
                    +k^2 \frac{m_e^2}{M^2} B_{\mathbf{k}} e^{ i m_e \mathbf{k} \cdot \mathbf{r} /M }  .
\end{align}
Plugging in, we get
\begin{align}
    \left( \frac{\hbar^2}{2\mu} k^2  \frac{m_h^2}{M^2} +  \hbar \omega_{{\rm l}, k}  + \frac{\hbar^2 k^2}{2M}
    \right) A_{\mathbf{k}} e^{-i m_h \mathbf{k} \cdot \mathbf{r} /M }
    - \left( \frac{\hbar^2}{2\mu} k^2  \frac{m_e^2}{M^2} +  \hbar \omega_{{\rm l}, k}  + \frac{\hbar^2 k^2}{2M}
    \right) B_{\mathbf{k}} e^{ i m_e \mathbf{k} \cdot \mathbf{r} /M }
    = V_k^* \left(e^{-i m_h \mathbf{k}\cdot\mathbf{r} /M} - e^{i m_e \mathbf{k}\cdot\mathbf{r} /M} \right)
\end{align}
One has
\begin{align}
    \frac{\hbar^2}{2\mu} k^2  \frac{m_h^2}{M^2}  + \frac{\hbar^2 k^2}{2M}
    &= \frac{\hbar^2 k^2}{2M} \left( \frac{m_h^2 M}{m_e m_h M } +1 \right)
     = \frac{\hbar^2 k^2}{2M} \frac{m_h + m_e}{m_e} = \frac{\hbar^2 k^2}{2m_e}, \\
    \frac{\hbar^2}{2\mu} k^2  \frac{m_e^2}{M^2}  + \frac{\hbar^2 k^2}{2M}
    &= \frac{\hbar^2 k^2}{2M} \left( \frac{m_e^2 M}{m_e m_h M } +1 \right)
     = \frac{\hbar^2 k^2}{2M} \frac{m_e + m_h}{m_h} = \frac{\hbar^2 k^2}{2m_h},
\end{align}
so that
\begin{align}
    A_{\mathbf{k}} = \frac{V_k^*}{\left(\hbar \omega_{{\rm l}, k}  + \frac{\hbar^2 k^2}{2m_e}
    \right)}, \quad
    B_{\mathbf{k}} = \frac{V_k^*}{\left(  \hbar \omega_{{\rm l}, k}  + \frac{\hbar^2 k^2}{2m_h}
    \right)}.
\end{align}
Plugging into the energy functional, we reach at
\begin{align}
    W_{\rm X} &= \sum_k \frac{\hbar^2 k^2}{2M} |V_k|^2 
    \left( \frac{m_h}{m_e \left(  \hbar \omega_{{\rm l}, k}  + \frac{\hbar^2 k^2}{2m_e} \right)^2}
    + \frac{m_e}{m_h \left(  \hbar \omega_{{\rm l}, k}  + \frac{\hbar^2 k^2}{2m_h} \right)^2}
    + \frac{2 \cos (\mathbf{k} \cdot \mathbf{r}) }{\left(  \hbar \omega_{{\rm l}, k}  + \frac{\hbar^2 k^2}{2m_e} \right)
    \left(  \hbar \omega_{{\rm l}, k}  + \frac{\hbar^2 k^2}{2m_h} \right)}
    \right) \notag \\
    &+\sum_k \left( \hbar \omega_{{\rm l}, k}  + \frac{\hbar^2 k^2}{2M}
    \right) |V_k|^2
    \left(\frac{1}{ \left(  \hbar \omega_{{\rm l}, k}  + \frac{\hbar^2 k^2}{2m_e} \right)^2}
    + \frac{1}{ \left(  \hbar \omega_{{\rm l}, k}  + \frac{\hbar^2 k^2}{2m_h} \right)^2}
    - \frac{2 \cos (\mathbf{k} \cdot \mathbf{r}) }{\left(  \hbar \omega_{{\rm l}, k}  + \frac{\hbar^2 k^2}{2m_e} \right)
    \left(  \hbar \omega_{{\rm l}, k}  + \frac{\hbar^2 k^2}{2m_h} \right)}
    \right) \notag\\
    &-\sum_k 2|V_k|^2
    \left( \frac{1 -\cos (\mathbf{k}\cdot\mathbf{r})}{  \hbar \omega_{{\rm l}, k}  + \frac{\hbar^2 k^2}{2m_e} }
    + \frac{1 -\cos (\mathbf{k}\cdot\mathbf{r})}{  \hbar \omega_{{\rm l}, k}  + \frac{\hbar^2 k^2}{2m_h} }
    \right)
    \notag \\
    &=-\sum_k  |V_k|^2 
    \left( \frac{1}{  \hbar \omega_{{\rm l}, k}  + \frac{\hbar^2 k^2}{2m_e} }
    + \frac{1}{\hbar \omega_{{\rm l}, k}  + \frac{\hbar^2 k^2}{2m_h} } \right) \notag\\
    &+\sum_k 2|V_k|^2 \cos (\mathbf{k}\cdot\mathbf{r})
    \left( -\frac{\hbar \omega_{{\rm l}, k}}{\left(  \hbar \omega_{{\rm l}, k}  + \frac{\hbar^2 k^2}{2m_e} \right)
    \left(  \hbar \omega_{{\rm l}, k}  + \frac{\hbar^2 k^2}{2m_h} \right)}
    +\frac{1}{  \hbar \omega_{{\rm l}, k}  + \frac{\hbar^2 k^2}{2m_e} }
    +\frac{1}{  \hbar \omega_{{\rm l}, k}  + \frac{\hbar^2 k^2}{2m_h} }\right) \notag \\
    &=-\sum_k  |V_k|^2 
    \left( \frac{1}{  \hbar \omega_{{\rm l}, k}  + \frac{\hbar^2 k^2}{2m_e} }
    + \frac{1}{\hbar \omega_{{\rm l}, k}  + \frac{\hbar^2 k^2}{2m_h} } \right) \notag\\
    &+\sum_k |V_k|^2 \cos (\mathbf{k}\cdot\mathbf{r})
    \left( \frac{m_e + m_h}{m_e-m_h} 
    \left( \frac{1}{  \hbar \omega_{{\rm l}, k}  + \frac{\hbar^2 k^2}{2m_e} }
    -\frac{1}{  \hbar \omega_{{\rm l}, k}  + \frac{\hbar^2 k^2}{2m_h} } \right)
    +\frac{1}{  \hbar \omega_{{\rm l}, k}  + \frac{\hbar^2 k^2}{2m_e} }
    +\frac{1}{  \hbar \omega_{{\rm l}, k}  + \frac{\hbar^2 k^2}{2m_h} }\right) \notag \\
    &=-\sum_k  |V_k|^2 
    \left( \frac{1}{  \hbar \omega_{{\rm l}, k}  + \frac{\hbar^2 k^2}{2m_e} }
    + \frac{1}{\hbar \omega_{{\rm l}, k}  + \frac{\hbar^2 k^2}{2m_h} } \right) 
    +\sum_k 2 |V_k|^2 \frac{\cos (\mathbf{k}\cdot\mathbf{r})}{m_h-m_e}
    \left( \frac{m_h}{  \hbar \omega_{{\rm l}, k}  + \frac{\hbar^2 k^2}{2m_h} }
    - \frac{m_e}{  \hbar \omega_{{\rm l}, k}  + \frac{\hbar^2 k^2}{2m_e} }\right).
\end{align}
The first term corresponds to single polaron binding energies of electron and hole.
The second term results in the re-scaled Coulomb interaction, and is evaluated as
\begin{align}
    &\frac{A}{(2\pi)^2}
    \frac{e^2  \omega_{\rm t} \hbar (r_0 - r_\infty) }{A (4 \varepsilon_0)}
    \int\limits_0^\infty \int\limits_0^{2\pi} 
     \frac{2\cos( k r \cos \varphi) }{ \sqrt{ (\varepsilon +  r_0 k ) (\varepsilon +  r_\infty k )^3} }  
      \left( \frac{m_h}{  \hbar \omega_{\rm t} \sqrt{\frac{\varepsilon+r_0 k}{\varepsilon+r_\infty k}}  + \frac{\hbar^2 k^2}{2m_h} }
    - \frac{m_e}{  \hbar \omega_{\rm t} \sqrt{\frac{\varepsilon+r_0 k}{\varepsilon+r_\infty k}}  + \frac{\hbar^2 k^2}{2m_e} }\right)
     \frac{ k{\rm d} k {\rm d}\varphi  }{m_h-m_e} \notag \\
     &=\frac{1}{(4\pi)}
    \frac{e^2  \omega_{\rm t} \hbar (r_0 - r_\infty) }{ 4 \pi \varepsilon_0}
    \int\limits_0^\infty  
     \frac{4\pi J_0 ( k r) }{ \sqrt{ (\varepsilon +  r_0 k ) (\varepsilon +  r_\infty k )^3} }  \frac{\sqrt{\varepsilon+r_\infty k}}{\hbar \omega_{\rm t} \sqrt{\varepsilon+r_0 k}  }
      \left( \frac{m_h}{ 1  + \frac{\hbar k^2}{2m_h\omega_{\rm t}} \sqrt{\frac{\varepsilon+r_\infty k}{\varepsilon+r_0 k}} }
    - \frac{m_e}{ 1  + \frac{\hbar k^2}{2m_e \omega_{\rm t}} \sqrt{\frac{\varepsilon+r_\infty k}{\varepsilon+r_0 k}} }\right)
     \frac{ k{\rm d} k  }{m_h-m_e} \notag \\
     &=    \frac{e^2   (r_0 - r_\infty) }{ 4 \pi  \varepsilon_0}
    \int\limits_0^\infty  
     \frac{ J_0 ( k r) }{ (\varepsilon +  r_0 k ) (\varepsilon +  r_\infty k ) }  
      \left( \frac{m_h}{ 1  + r_{h,{\rm t}}^2 k^2 \sqrt{\frac{\varepsilon+r_\infty k}{\varepsilon+r_0 k}} }
    - \frac{m_e}{ 1  + r_{e,{\rm t}}^2 k^2 \sqrt{\frac{\varepsilon+r_\infty k}{\varepsilon+r_0 k}} }\right)
     \frac{ k{\rm d} k  }{m_h-m_e} \notag \\
     &=    \frac{e^2 }{ 4 \pi \varepsilon_0 r_\infty}
     \frac{\sigma_0-1}{2}
    \int\limits_0^\infty  
     \frac{ 2 J_0 ( \Tilde{k} \varepsilon r/r_\infty) }{ (1 +  \sigma_0 \Tilde{k} ) (1 + \Tilde{k} ) }  
      \left( \frac{m_h}{ 1  + \sigma_{h,{\rm t}}^2 \Tilde{k}^2  \sqrt{\frac{1+\Tilde{k}}{1+\sigma_0 \Tilde{k}}} }
    - \frac{m_e}{ 1  + \sigma_{e,{\rm t}}^2 \Tilde{k}^2  \sqrt{\frac{1+\Tilde{k}}{1+\sigma_0 \Tilde{k}}} }\right)
     \frac{ \Tilde{k}{\rm d} \Tilde{k}  }{m_h-m_e} 
\end{align}
Therefore, the effective Coulomb interaction is
\begin{align}
     V_{\rm eff} (r) &= V_{\rm KR} (r,r_\infty) + \sum_k 2 |V_k|^2 \frac{\cos (\mathbf{k}\cdot\mathbf{r})}{\Delta m}
    \left( \frac{m_h}{  \hbar \omega_{{\rm l}, k}  + \frac{\hbar^2 k^2}{2m_h} }
    - \frac{m_e}{  \hbar \omega_{{\rm l}, k}  + \frac{\hbar^2 k^2}{2m_e} }\right) \notag \\
    &= \boxed{ V_{\rm KR} (r,r_\infty) + 
    \frac{e^2 }{ 4 \pi \varepsilon_0 r_\infty}
     \frac{\sigma_0-1}{2}
    \int\limits_0^\infty  
     \frac{ 2 J_0 ( \Tilde{k} \varepsilon r/r_\infty) }{ (1 +  \sigma_0 \Tilde{k} ) (1 + \Tilde{k} ) }  
      \left( \frac{m_h}{ 1  + \sigma_{h,{\rm t}}^2 \Tilde{k}^2  \sqrt{\frac{1+\Tilde{k}}{1+\sigma_0 \Tilde{k}}} }
    - \frac{m_e}{ 1  + \sigma_{e,{\rm t}}^2 \Tilde{k}^2  \sqrt{\frac{1+\Tilde{k}}{1+\sigma_0 \Tilde{k}}} }\right)
     \frac{ \Tilde{k}{\rm d} \Tilde{k}  }{\Delta m}},
\end{align}
where $\Delta m =m_h-m_e$.
We then apply an approximation $\sqrt{(1+\Tilde{k})/(1+\sigma_0\Tilde{k})} \approx 1/\sqrt{\sigma_0}$, and factorize the integrand:
\begin{align}
    \frac{2\Tilde{k}}{ (1 +  \sigma_0 \Tilde{k} ) (1 + \Tilde{k} )} \frac{1}{1+ \epsilon_i \Tilde{k}^2} 
    = \frac{2}{(k+1) (\sigma_0-1) (\epsilon_i +1)}
    -\frac{2 \sigma_0^2}{(\sigma_0-1) (k \sigma_0+1) \left(\sigma_0^2+\epsilon_i \right)}
    +\frac{2 \left(k \epsilon_i ^2-k \sigma_0 \epsilon_i +\sigma_0 \epsilon_i +\epsilon_i \right)}{(\epsilon_i +1) \left(k^2 \epsilon_i +1\right) \left(\sigma_0^2+\epsilon_i \right)}
\end{align}
where $\epsilon_i = \sigma_{i,{\rm t}}^2/\sqrt{\sigma_0}$, $i=e,h$.
Evaluating the terms one by one, we get
\begin{align}
    \int\limits_0^\infty \frac{2 J_0 ( \Tilde{k} \varepsilon r/r_\infty)}{(k+1) (\sigma_0-1) (\epsilon_i +1)} {\rm d} \Tilde{k}
    &= \frac{\pi}{(\sigma_0-1) (\epsilon_i +1)} 
    \left[ H_0 \left(\frac{\varepsilon r}{r_\infty} \right) - Y_0 \left(\frac{\varepsilon r}{r_\infty} \right)  \right], \\ 
    \int\limits_0^\infty \frac{2 \sigma_0^2 J_0 ( \Tilde{k} \varepsilon r/r_\infty) }{(\sigma_0-1) (k \sigma_0+1) \left(\sigma_0^2+\epsilon_i \right)}  {\rm d} \Tilde{k} 
    &=  \frac{\pi \sigma_0}{(\sigma_0-1) (\epsilon_i +\sigma_0^2)} 
    \left[ H_0 \left(\frac{\varepsilon r}{r_\infty \sigma_0} \right) - Y_0 \left(\frac{\varepsilon r}{r_\infty \sigma_0} \right)  \right],  \\
    \int\limits_0^\infty J_0 ( \Tilde{k} \varepsilon r/r_\infty) \frac{2 \epsilon_i \left(k \epsilon_i -k \sigma_0  +\sigma_0  +1 \right)}{(\epsilon_i +1) \left(k^2 \epsilon_i +1\right) \left(\sigma_0^2+\epsilon_i \right)} {\rm d} \Tilde{k} 
    &= \frac{1}{(1+\epsilon_i) (\epsilon_i +\sigma_0^2)}
    \left( \pi (1+ \sigma_0) \sqrt{\epsilon_i} \left[ I_0 \left(\frac{\varepsilon r}{r_\infty \sqrt{\epsilon_i}} \right) 
    - L_0 \left(\frac{\varepsilon r}{r_\infty \sqrt{\epsilon_i}} \right)  \right] \right. \notag \\
    & \left. + 2 (\epsilon_i - \sigma_0) K_0 \left(\frac{\varepsilon r}{r_\infty \sqrt{\epsilon_i}} \right) \right),
\end{align}
where $I_0$, $K_0$, are modified Bessel functions, and $L_0$ is the Struve function.
Hence, the overall potential is
\begin{align}
     V_{\rm eff} (r) &= V_{\rm KR} (r,r_\infty) + \frac{e^2 }{ 4 \pi \varepsilon_0 r_\infty} 
     \left( i_1 + i_2 + i_3 \right),
\end{align}
where
\begin{align}
    i_1 &= \frac{\pi}{2} \left[ \frac{m_h}{\Delta_m (\epsilon_h +1)}
    -\frac{m_e}{\Delta_m (\epsilon_e +1)} \right] 
    \left[ H_0 \left(\frac{\varepsilon r}{r_\infty} \right) - Y_0 \left(\frac{\varepsilon r}{r_\infty} \right)  \right], \\
    i_2 &= - \frac{\pi}{2} \left[ \frac{\sigma_0 m_h}{\Delta_m (\epsilon_h +\sigma_0^2)}
    -\frac{\sigma_0 m_e}{\Delta_m (\epsilon_e +\sigma_0^2)} \right] 
    \left[ H_0 \left(\frac{\varepsilon r}{r_0} \right) - Y_0 \left(\frac{\varepsilon r}{r_0} \right)  \right], \\
    i_3 &= \frac{\sigma_0-1}{2} \frac{m_h}{\Delta m}
    \frac{1}{(1+\epsilon_h) (\epsilon_h +\sigma_0^2)}
    \left( \pi (1+ \sigma_0) \sqrt{\epsilon_h} \left[ I_0 \left(\frac{\varepsilon r}{r_\infty \sqrt{\epsilon_h}} \right) 
    - L_0 \left(\frac{\varepsilon r}{r_\infty \sqrt{\epsilon_h}} \right)  \right]
    + 2 (\epsilon_h - \sigma_0) K_0 \left(\frac{\varepsilon r}{r_\infty \sqrt{\epsilon_h}} \right) \right) \notag \\
    &- \frac{\sigma_0-1}{2} \frac{m_e}{\Delta m}
    \frac{1}{(1+\epsilon_e) (\epsilon_e +\sigma_0^2)}
    \left( \pi (1+ \sigma_0) \sqrt{\epsilon_e} \left[ I_0 \left(\frac{\varepsilon r}{r_\infty \sqrt{\epsilon_e}} \right) 
    - L_0 \left(\frac{\varepsilon r}{r_\infty \sqrt{\epsilon_e}} \right)  \right]
    + 2 (\epsilon_e - \sigma_0) K_0 \left(\frac{\varepsilon r}{r_\infty \sqrt{\epsilon_e}} \right) \right) 
\end{align}
We note that $\epsilon_i \ll 1$, and expand to first order, yielding in 
\begin{align}
    i_1 &\approx \frac{\pi}{2}  \frac{m_h -m_e - m_h \epsilon_h + m_e \epsilon_e  }{\Delta_m }
    \left[ H_0 \left(\frac{\varepsilon r}{r_\infty} \right) - Y_0 \left(\frac{\varepsilon r}{r_\infty} \right)  \right]
    = \frac{\pi}{2}  \left[ H_0 \left(\frac{\varepsilon r}{r_\infty} \right) - Y_0 \left(\frac{\varepsilon r}{r_\infty} \right)  \right], \\
    i_2 & \approx - \frac{\pi}{2} 
    \frac{m_h -m_e - (m_h \epsilon_h -m_e \epsilon_e) / \sigma_0^2}{\Delta_m \sigma_0 }
    \left[ H_0 \left(\frac{\varepsilon r}{r_0} \right) - Y_0 \left(\frac{\varepsilon r}{r_0} \right)  \right]
    = -\frac{\pi}{2\sigma_0} 
    \left[ H_0 \left(\frac{\varepsilon r}{r_0} \right) - Y_0 \left(\frac{\varepsilon r}{r_0} \right)  \right],
\end{align}
where we use
\begin{align}
    m_h \epsilon_h - m_e \epsilon_e 
    = (m_h \sigma_{h,{\rm t}}^2 - m_e \sigma_{e,{\rm t}}^2) /\sqrt{\sigma_0} =0.
\end{align}
One can now note the cancellation
\begin{align}
    V_{\rm KR} (r,r_\infty) + \frac{e^2 }{ 4 \pi  \varepsilon_0 r_\infty} i_1 =0,
\end{align}
which eventually results in the following effective potential
\begin{align}
    V_{\rm eff} (r) 
    &\approx V_{\rm KR} (r,r_0)
    + \frac{e^2 }{ 4 \pi \varepsilon_0 r_\infty}
     \frac{\sigma_0-1}{2} \left[ \frac{m_h}{\Delta m}
    \frac{\pi (1+\sigma_0) \Tilde{\sigma}_{h,{\rm t}} 
    \left( I_0\left( \frac{\varepsilon r}{ \Tilde{\sigma}_{h,{\rm t}} r_\infty} \right)
    -L_0\left( \frac{\varepsilon r}{ \Tilde{\sigma}_{h,{\rm t}} r_\infty} \right) \right) 
    + 2 (\Tilde{\sigma}_{h,{\rm t}}^2-\sigma_0) K_0\left( \frac{\varepsilon r}{ \Tilde{\sigma}_{h,{\rm t}} r_\infty} \right)}
    {(1+\Tilde{\sigma}_{h,{\rm t}}^2) (\sigma_0^2+\Tilde{\sigma}_{h,{\rm t}}^2) }
    \right. \notag \\
    & \hspace{85pt}   \left. -\frac{m_e}{\Delta m}
    \frac{\pi (1+\sigma_0) \Tilde{\sigma}_{e,{\rm t}} 
    \left( I_0\left(\frac{\varepsilon r}{\Tilde{\sigma}_{e,{\rm t}} r_\infty} \right)
    -L_0\left( \frac{\varepsilon r}{ \Tilde{\sigma}_{e,{\rm t}} r_\infty} \right) \right) 
    + 2 (\Tilde{\sigma}_{e,{\rm t}}^2-\sigma_0) K_0\left( \frac{\varepsilon r}{ \Tilde{\sigma}_{e,{\rm t}} r_\infty} \right)}
    {(1+\Tilde{\sigma}_{e,{\rm t}}^2) (\sigma_0^2+\Tilde{\sigma}_{e,{\rm t}}^2) }
    \right],
\end{align}
where we recall $\Tilde{\sigma}_{i, {\rm t}} = \sigma_{i, {\rm t}} /\sigma_0^{1/4}$, $i=e,h$.
We can further simplify to the form
\begin{align}
    V_{\rm eff} (r) 
    &\approx V_{\rm KR} (r,r_0)
    + E_0
    \left[ \frac{m_h}{\Delta m}
    \Phi_{\Tilde{\sigma}_{h,{\rm t}} } (\varepsilon r/r_\infty)
    -\frac{m_e}{\Delta m}
    \Phi_{\Tilde{\sigma}_{e,{\rm t}} } (\varepsilon r/r_\infty)
    \right],
\end{align}
where
\begin{align}
    \Phi_{\nu}( x ) =
    \dfrac{\sigma_0-1 }
    {2(1+\nu^2) (\sigma_0^2+\nu^2) }
    \left[ (1+\sigma_0) \nu\pi 
    \left[ I_0\left( x/\nu\right)
    -L_0\left( x/\nu\right) \right] 
    + 2 (\nu^2-\sigma_0) K_0\left(x/\nu\right) \right].
\end{align}

\end{document}